\documentclass[journal]{IEEEtran}

\usepackage{cite}
\usepackage[cmex10]{amsmath}

\DeclareMathOperator{\Tr}{Tr}

\DeclareMathSizes{10}{9}{7}{6}
\newcommand{\vect}{\operatorname{vec}}

\usepackage{url}

\usepackage{amssymb}   
\usepackage{amsxtra}
\usepackage{amscd}
\usepackage{amsthm}
\usepackage{amsxtra}
\usepackage{amscd}
\usepackage{amsthm}
\usepackage{tikz}
\usetikzlibrary{matrix}
\usepackage{mathtools}
\usepackage{graphicx}   
\usepackage{wrapfig}
\usepackage{layouts}
\usepackage{booktabs,makecell,multirow,tabularx}
\usepackage{arydshln}
\usepackage{flushend}
\usepackage{xparse}
\usepackage[caption=false]{subfig}
\NewDocumentCommand{\overarrow}{O{=} O{\uparrow} m}{%
	\overset{\makebox[0pt]{\begin{tabular}{@{}c@{}}#3\\[0pt]\ensuremath{#2}\end{tabular}}}{#1}
}
\NewDocumentCommand{\underarrow}{O{=} O{\downarrow} m}{%
	\underset{\makebox[0pt]{\begin{tabular}{@{}c@{}}\ensuremath{#2}\\[0pt]#3\end{tabular}}}{#1}
}

\usepackage{verbatim}
\usepackage{stfloats}
\usepackage{blindtext}
\usepackage{abraces}
\usepackage{algpseudocode}
\usepackage{algorithm}
\usepackage{color}

\usepackage{xparse}
\renewcommand{\thefootnote}{\arabic{footnote}}

\usepackage{balance}

\usetikzlibrary{decorations.pathreplacing,
	matrix,
	positioning,
}

\usepackage[bottom]{footmisc}
\usepackage{tikz}

\usetikzlibrary{decorations.pathreplacing}
\newcommand{\vast}{\bBigg@{4}}
\newcommand{\Vast}{\bBigg@{5}}
\usepackage{float}
\usepackage{xcolor,cite,etoolbox}
\usepackage[numbers,sort&compress]{natbib}

\usepackage{filecontents}
\renewcommand{\arraystretch}{2}

\bstctlcite{IEEEexample:BSTcontrol}
\begin{document}
\bstctlcite{IEEEexample:BSTcontrol}

\author{Zehra~Yigit,~\IEEEmembership{Member,~IEEE,} Sefa~Kayraklik,~\IEEEmembership{Graduate Student Member,~IEEE,} 
		Ertugrul~Basar,~\IEEEmembership{Fellow,~IEEE,}
        Ali~Gorcin,~\IEEEmembership{Senior Member,~IEEE}     
        \thanks{Z.Yigit is with Artificial Intelligence and 6G Laboratory (6GEN. LAB.), Turkcell Iletisim Hizmetleri Inc., Istanbul, Turkiye. E-mail: zehra.yigit@turkcell.com.tr}
        \thanks{S. Kayraklik is with the Communications and Signal Processing Research (HİSAR) Lab., TÜBİTAK-BİLGEM, Kocaeli, Turkiye, and also with the Department of Electrical and Electronics Engineering, Koc University, Sariyer, Istanbul, Turkiye. E-mail: sefa.kayraklik@tubitak.gov.tr}
        \thanks{E. Basar is with the Department of Electrical Engineering, Tampere University, Tampere, Finland, on leave from the Department of Electrical and Electronics Engineering, Koc University, Sariyer, Istanbul, Turkiye. E-mail: ertugrul.basar@tuni.fi and ebasar@ku.edu.tr.}
        \thanks{A. Gorcin is with the Communications and Signal Processing Research (HİSAR) Lab., TÜBİTAK-BİLGEM, Kocaeli, Turkiye, and also with the Electronics and Communication Department, Istanbul Technical University, Istanbul, Turkiye. E-mail: aligorcin@itu.edu.tr}
}

\title{Dual Target-Mounted RISs-Assisted ISAC Against Eavesdropping and Malicious Interference}

\markboth{Journal of \LaTeX\ Class Files,~Vol.~X, No.~X, Mar~2026}%
{ \MakeLowercase{\textit{et al.}}: PLS for RIS-assisted ISAC Systems}

\maketitle

\begin{abstract}
The synergy between integrated sensing and communication (ISAC) and reconfigurable intelligent surfaces (RISs) unlocks novel applications and advanced services for next-generation wireless networks, yet also introduces new security challenges.
In this study, a novel dual target-mounted RISs-assisted ISAC scheme is proposed, where a base station with ISAC capability performs sensing of two unmanned aerial vehicle (UAV) targets, one of which is legitimate and the other is eavesdropper, while communicating with the users through an RIS mounted on the legitimate UAV target. The proposed scheme addresses dual security threats posed by a hostile UAV target: eavesdropping on legitimate user communications and random interference attacks launched by a malicious RIS mounted on this eavesdropper UAV target, aiming to disrupt secure transmissions. {Moreover, malicious RIS interference is also optimized for a worst-case scenario, in which both the channel state information (CSI) and the transmit beamforming of the base station are assumed to be fully compromised by a malicious RIS-mounted eavesdropper UAV.  } A non-convex optimization problem maximizing the secrecy rate of the users is formulated, and a semi-definite relaxation (SDR)-based two-stage solution is developed to optimize the transmit beamforming matrix of the base station and the phase shift coefficients of the legitimate RIS.  Extensive computer simulations are conducted to evaluate the robustness of the proposed solution under various system configurations. The proposed system's communication performance is assessed using the secrecy rate metric, while the sensing performance is evaluated through the signal-to-interference-plus-noise ratio and the Cramer–Rao bound (CRB) for angle-of-departure (AoD) estimation of the eavesdropper UAV target.
\end{abstract}

	\let\thefootnote\relax\footnote{This work was supported by The Scientific and Technological Research Council of Turkiye (TUBITAK) through the 1515 Frontier Research and Development Laboratories Support Program under Project 5229901 - 6GEN. Lab: 6G and Artificial Intelligence Laboratory, and also Grant 120E401.} 

\section{Introduction}

The paradigm shift from supporting communication solely between people to seamless interactions between people and connected devices has laid a foundation for a new generation demanding enhanced data rates, ultra-low latency, and massive device connectivity \cite{ghosh20195g}. As the sixth-generation (6G) era approaches, the foundational framework of its evolution has begun to shape through International Telecommunication Union (ITU) recommendations \cite{ITU-R_IMT-2030} and ongoing 3rd Generation Partnership Project (3GPP) releases \cite{3GPP_TR_38_914}. Accordingly, while building upon enhanced fifth-generation capabilities, 6G introduces three novel transformative usage scenarios: artificial intelligence-driven communication, ubiquitous connectivity, and integrated sensing and communication (ISAC), supporting diverse emerging applications from autonomous systems, satellite-connected networks, and advanced environmental perception \cite{ITU-R_IMT-2030}.

In 6G networks, ISAC technology offers enabling communication and environmental sensing in a unified hardware platform, exploiting shared frequency spectrum, resources, and signal processing capabilities \cite{liu2022integrated}. This opens up a wide range of applications across various environment-aware scenarios, including environmental monitoring, smart cities, autonomous transportation, and internet of things (IoT) \cite{liu2022integrated}. These applications leverage newly introduced advanced sensing-related capabilities such as high-precision mapping, sub-centimeter resolution imaging, and centimeter-level localization enabled by  ISAC technologies in 6G networks \cite{ITU-R_IMT-2030}.  

Although ISAC is now widely recognized as a key 6G technology, its earliest studies trace back to several decades, exploring the dual use of radar and communication systems \cite{liu2022integrated}. Since then,  this research has appeared under various terminologies, such as joint radar and communication (JRC) \cite{liu2020joint}, joint communication and sensing (JCAS) \cite{zhang2021enabling}, and dual-function radar-communication (DFRC) \cite{liu2018toward}, ultimately converging towards the concept of ISAC \cite{liu2022integrated}.

Another promising candidate technology in 6G networks is reconfigurable intelligent surfaces (RISs)-empowered communication,  offering a controllable wireless environment in a cost-effective manner \cite{di2020smart}. RISs are fundamentally based on meta-surface technology employing tunable electronic components such as PIN diodes or varactor diodes to dynamically control electromagnetic (EM) properties of incident signals \cite{basar2019wireless, di2020smart}. By switching between different states of these elements, RISs can manipulate EM characteristics of incident waves such as phase, amplitude, and polarization \cite{di2020smart}. This unprecedented capability of RISs positions them as a strong candidate technology for 6G wireless networks \cite{basar2019wireless}. Over the past few years,  research on RIS has primarily concentrated on improved signal quality \cite{basar2019wireless}, physical layer security \cite{kayraklik2024indoor}, enhanced capacity achievement \cite{zhang2020capacity}, and passive beamforming designs \cite{wu2019intelligent}, which leverages RIS to shape the propagation environment in a constructive manner. However, a few studies have also examined RIS as a potential security threat, demonstrating how maliciously configured surfaces can degrade a legitimate wireless system \cite{alexandropoulos2023counteracting, rivetti2024malicious}. 

As 6G studies advance, a growing number of studies for RIS-assisted ISAC schemes have emerged in the literature \cite{elbir2022rise}. In \cite{wang2023stars, zhang2017matrix}, the sensing-capable RIS concepts have been introduced to tackle the inherent severe path attenuation of RIS-aided system designs. In \cite{luo2023ris, zhang2024intelligent}, a joint beamforming and phase optimization algorithm has been developed to satisfy quality of service (QoS) requirements of both communication users and sensing targets. Moreover, target-mounted aerial RIS-aided ISAC schemes have been presented to improve coverage \cite{xu2024irs} and facilitate accurate localization  \cite{cho2024enhancing}, {improve air to ground communication \cite{saikia2025hybrid}, mitigate malicious jamming \cite{chen2025energy} and enhance target detection probability \cite{mizmizi2023target}.}  Furthermore, several studies have explored secure communication in RIS-assisted ISAC systems, considering scenarios where sensing targets could act as potential eavesdroppers intercepting legitimate users, while RISs are strategically configured to overcome security challenges effectively  \cite{yang2024secure, liu2023securing}. Although prior studies have investigated the effects of eavesdropper targets in RIS-aided ISAC schemes \cite{xu2024irs, yang2024secure, liu2023securing,  benchmark, shao2024target}, the joint impact of eavesdropper targets and malicious RIS attacks within an ISAC system has not yet been investigated. {This critical gap motivates this study, to jointly tackle these dual threats in an ISAC system. The optimization is performed under imperfect channel state information (CSI) of the eavesdropper channel, reflecting  a practically relevant scenario where the legitimate base station (BS) has incomplete and uncertain knowledge of the eavesdropper, while a malicious RIS simultaneously attempts to degrade secure transmissions through deliberate interference attacks.}

{ \subsection{Motivation and Contributions}

In \cite{saikia2025hybrid, chen2025energy}, the UAV-mounted RIS is employed merely as a supporting component to strengthen links for distant communication users and sensing targets, thereby improving both communication reliability and sensing accuracy within the ISAC system. Moreover, in \cite{mizmizi2023target}, sensing vehicle targets are lodged with RISs, each equipped with amplitude and phase controllable elements to deliberately back-reflect the probing signal and enhance its detectability. In the proposed dual target-mounted RIS-assisted ISAC schemes, unlike \cite{saikia2025hybrid, chen2025energy}, which integrate RISs on external mobile vehicles to enhance the communication and sensing performance of distant users and targets, the RISs are instead mounted directly on the detectable sensing targets. However, in contrast to \cite{mizmizi2023target}, which deploys an RIS to avoid treating the vehicle as an external target, our scheme diverges by strategically employing dual target-mounted RISs with opposing functions. While one RIS intentionally generates a malicious reflective response for legitimate users, the other is dedicated to enhancing physical layer security for those same users.    

Existing RIS-assisted secure ISAC systems \cite{xu2024irs, yang2024secure, liu2023securing, benchmark} simply treat the sensing target as a potential eavesdropper, and an external RIS is configured solely to enhance the secrecy performance of legitimate users. In contrast, the proposed scheme addresses a far more challenging problem when one of the targets is an eavesdropper equipped with a malicious RIS that introduces unpredictable interference into the legitimate system.  Although the BS and this legitimate RIS are completely unaware of the presence or behavior of the malicious RIS, the legitimate RIS is jointly optimized with the BS to reinforce the sensing and communication paths. }

In this paper, to overcome the above challenges, a novel RIS-assisted ISAC system is proposed, where a BS performs sensing of two unmanned aerial vehicle (UAV) targets while simultaneously establishing reliable communication links with the users through an RIS mounted on the legitimate UAV. The system model considers security threats posed by an eavesdropper UAV target that intercepts wireless communication signals while a malicious RIS, mounted on the eavesdropper UAV, attempts to disrupt the users' communication by launching  interference attacks. 
The major contributions of this study can be summarized as follows: 
\begin{itemize}
    \item A dual target-mounted RISs-assisted secure ISAC system is considered, where one UAV target both acts as an eavesdropper to intercept the communication between the BS and the users and launches interference attacks using the malicious RIS to disrupt the user communication links.
    \item {The malicious RIS interference is assumed to  encompass both random attacks and  a  worst-case scenario, in which the malicious RIS-mounted UAV is assumed to have perfect knowledge of the CSI and the transmit beamforming matrix of the legitimate BS.}   
    \item {
    To mitigate the eavesdropping and malicious RIS interference threats, an optimization problem is formulated under the assumption of imperfect eavesdropper CSI at the BS. The transmit beamforming matrix of the BS and the phase shift coefficients of a legitimate RIS mounted on a legitimate UAV are optimized to maximize the system secrecy rate while ensuring that the sensing performance requirements are satisfied.} To address this challenge, a semi-definite relaxation (SDR)-based two-stage solution is developed.
    \item As a communication performance metric, the sum secrecy rate of the legitimate users is investigated through extensive computer simulations. For sensing performance, beampattern gain, signal-to-interference-plus-noise ratio (SINR) of UAV targets is analyzed, and the Cramer-Rao bound (CRB) is derived to estimate the two-dimensional (2D) angle of departures (AoDs) of UAV targets. {Moreover, a worst-case malicious RIS interference scenario is investigated, where the malicious RIS-mounted UAV eavesdropper is assumed to have perfect knowledge of the transmit beamforming matrix of the BS and the CSI of both the BS and the legitimate users.}
    \item Through comprehensive simulation results, the effectiveness of the proposed algorithm is demonstrated, showing notable performance improvements in the secrecy rate of the communication users while simultaneously enhancing key sensing performance metrics, namely the SINR and the CRB. {Furthermore, comparisons against a benchmark scheme \cite{benchmark} are provided to further assess the performance and robustness of the proposed approach under more challenging and demanding threat conditions.
}
\end{itemize}



The rest of the paper is organized as follows. Section II introduces the proposed UAV-mounted RIS-assisted secure ISAC framework. In Section III, the problem formulation for beamforming design and legitimate RIS phase optimization is provided. Section IV provides the performance evaluation for both the communication and sensing subsystems, and Section V concludes the paper.

\textit{Notation:} Unless otherwise specified, scalars are denoted by italic letters (i.e., $x$), while vectors and matrices are represented by boldface lower letters (i.e., $\mathbf{x}$) and boldface upper letters (i.e., $\mathbf{X}$), respectively. $\mathrm{diag}(\mathbf{\mathbf{x}})$ represents a diagonal matrix whose diagonal elements are the elements of the vector $\mathbf{x}$, while $\mathbf{I}$ denotes the identity matrix.  $|x|$ and $\left\| \mathbf{X}\right\|_{\mathrm{F}}$ stand for the absolute value of a scalar and the Frobenius norm of a matrix, respectively. $x^*$ is the conjugate of  $x$, while $\mathbf{X}^{-1}$, $\mathbf{X}^\mathrm{T}$, and $\mathbf{X}^\mathrm{H}$ represent the inverse, transposition, and Hermitian of a matrix, respectively. $\vect(\cdot)$ stands for vectorization operator, while $\Tr(\cdot)$ represents trace operator. $\otimes$ and  $\odot$ stand for Kronecker and Hadamard products, respectively. $\mathfrak{Re}(\cdot)$ and $\mathfrak{Im}(\cdot)$ denote real and imaginary components of a complex number, respectively. $\mathcal{CN}(\varpi, \sigma^2)$ represents the distribution of a complex Gaussian random variable with mean $\varpi$ and variance $\sigma^2$. $\mathbb{C}^{M\times L}$ denotes the space of complex matrices  with dimensions of $M\times L$. $\mathcal{O}$ stands for  big-O notation and $\mathbf{X}\succeq0$ denotes positive semi-definiteness of matrix $\mathbf{X}$.

\section{System Model}
In this section, the system model of the proposed UAV-mounted RISs-assisted secure ISAC system is presented. 

\subsection{Dual UAV-mounted RISs-assisted ISAC System}
\begin{figure}[!t]
    \centering
    \includegraphics[width=1\linewidth]{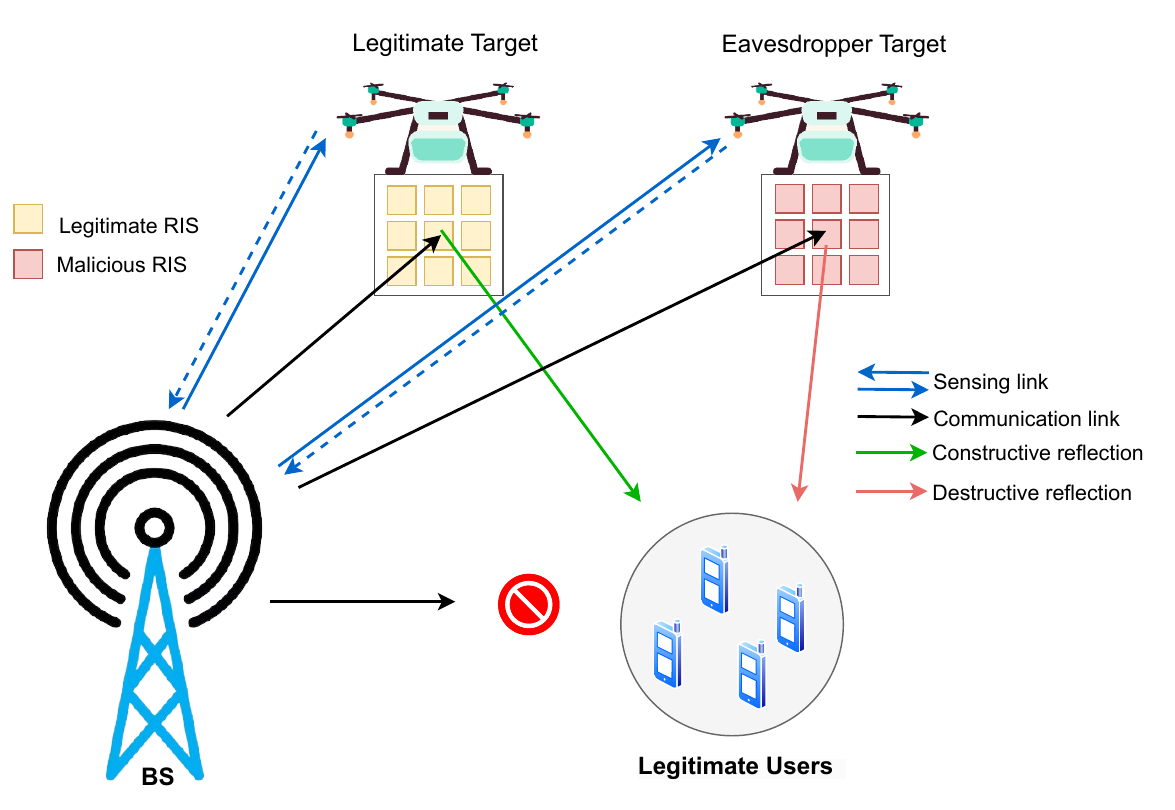}    
    \caption{Target-mounted RISs-assisted  ISAC scheme.}
    \label{Fig1}
\end{figure}

In the proposed scheme, as illustrated in Fig. \ref{Fig1}, a BS, equipped with $T_x$ transmit antennas {in a uniform linear array (ULA) structure to enable horizontal beam steering}  for sensing two UAV targets while simultaneously establishing communication links with $K$ single-antenna legitimate users via RISs mounted on those UAVs. Among two UAV targets, one is legitimate, while the other attempts to eavesdrop on the communication users. The legitimate UAV target carries a legitimate RIS to sustain reliable communication links between the BS and the users, whereas the eavesdropping UAV target is equipped with a malicious RIS, introducing random interference attacks to disrupt the communication links between the BS and the users. {Both the legitimate and malicious RISs are assumed to be passive planar surfaces, which naturally conform to a uniform planar array (UPA) configuration, comprising $N_L$ and 
$N_M$ reflecting elements, respectively, to enable precise phase shift control across both horizontal and vertical dimensions. }

The reflection matrices of legitimate and malicious RIS can be respectively given as 
\begin{align}
& \mathbf{\Omega}_L\in\mathbb{C}^{N_L\times N_L} = \mathrm{diag}\left( e^{j\psi_1}, e^{j\psi_2}, \cdots,e^{j\psi_{N_L}} \right),\label{eq.l-ris} \\
& \mathbf{\Omega}_M\in\mathbb{C}^{N_M\times N_M} = \mathrm{diag}\left( e^{j\zeta_1}, e^{j\zeta_2}, \cdots,e^{j\zeta_{N_M}} \right),\label{eq.m-ris} 
\end{align}
where  $\psi_l$ is the phase shift of $l$-th reflecting element of the legitimate RIS and $\zeta_m$ is  the phase shift of the $m$-th reflecting element of the malicious RIS, for $l\in\{1,\cdots,N_L\}$ and $m\in\{1,\cdots,N_M\}$. {Since the malicious RIS performs random interference attacks, }each $\zeta_m$ is assumed to be independently and identically distributed (i.i.d.) complex Gaussian random variable and follows $\sim\mathcal{CN}(0,1)$ distribution, while $\psi_l$ is optimized for enhancing secure transmission. 
In the proposed scheme, for enabling joint sensing and communication simultaneously,  the BS transmits a unified signal at the $t$-th time interval as follows
\begin{align}
 &\mathbf{x}(t)\in\mathbb{C}^{T_x\times 1} = \mathbf{W}_c{\mathbf{x}_c}(t) +    \mathbf{W}_s{\mathbf{x}_s}(t), \\
 &\hspace{1.95cm}= \bigl[
     \mathbf{W}_c,  \mathbf{W}_s
\bigr]\bigl[    {\mathbf{x}_c}(t), {\mathbf{x}_s}(t)\bigr]^\mathrm{T},
\end{align}
where $\mathbf{W}_c\in\mathbb{C}^{T_x\times K}$ and $\mathbf{W}_s\in\mathbb{C}^{T_x\times T_x}$  are the transmit beamforming matrix for 
communication users and sensing UAV targets, respectively. { The communication signal for the legitimate users is denoted by ${\mathbf{x}_c}(t)\in\mathbb{C}^{K\times 1}$ } and satisfying $\mathbb{E}\{ \mathbf{x}_c(t)\mathbf{x}_c(t)^{\mathrm{H}} \}=\mathbf{I}_K$, while and ${\mathbf{x}_s}(t)\in\mathbb{C}^{T_x\times 1}$ is frequency modulated continuous wave (FMCW) sensing signal \cite{xiao2022waveform} satisfying $\mathbb{E}\{ \mathbf{x}_s(t)\mathbf{x}_s(t)^{\mathrm{H}} \}=\mathbf{I}_{T_x}$ and also orthogonal to the communication signal to prevent mutual information  $\mathbb{E}\{\mathbf{x}_s(t)\mathbf{x}_c(t)^{\mathrm{H}} \}=\mathbf{0}_{T_x\times K}$. Therefore, for $\mathbf{W}\in\mathbb{C}^{T_x\times (T_x+K)}=\bigl[\mathbf{W}_c, \mathbf{W}_s\bigr]$ being the overall transmit beamforming matrix, the covariance of the transmit signal becomes
\begin{align}
& \mathbf{R_x} = \mathbb{E}\{ \mathbf{x}(t)\mathbf{x}(t)^{\mathrm{H}}\},\\
&\hspace{0.5cm}= \mathbf{W}\mathbf{W}^{\mathrm{H}}.
\end{align}
Therefore, the maximum total transmit power constraint at the BS can be calculated as
\begin{equation}
P_T\geq \mathrm{Tr}(\mathbf{R}_x).
\end{equation}


\subsection{Sensing  Model} \label{sec:sensing}
In the proposed scheme, both the legitimate and eavesdropper UAV targets are assumed to have line-of-sight (LoS) links with the BS. The array responses between the BS and the UAV targets are modeled using the 2D AoD, defined by the horizontal and vertical angles $\theta^{\text{BS}}_i$ and $\phi^{\text{BS}}_i$,  respectively, where $i\in\{L, E\}$  indicates legitimate UAV ($L$-UAV) and eavesdropper UAV ($E$-UAV), respectively. {  $\theta^{\text{BS}}_L$ and $\phi^{\text{BS}}_L$ of the $L$-UAV are perfectly known to the BS; however, $\theta^{\text{BS}}_E$ and the angle estimates of $\phi^{\text{BS}}_E$ related to $E$-UAV  are imperfect and mismatches may occur \cite{jia2023physical}. Therefore, these angles can be expressed as   $\phi^{\text{BS}}_E = \tilde{\phi}^{\text{BS}}_E + \Delta{\phi}_E$
 and $  \theta^{\text{BS}}_E = \tilde{\theta}^{\text{BS}}_E + \Delta{\theta}_E$. While $\tilde{\phi}^{\text{BS}}_E$ and $\tilde{\theta}^{\text{BS}}_E$ are actual horizontal and vertical angles, and $\Delta{\phi}_E\sim\mathcal{CN}(0, \sigma_{\phi}^2)$ and $\Delta{\theta}_E\sim\mathcal{CN}(0, \sigma_{\theta}^2)$ are the horizontal and vertical angle errors, respectively, 
where $\sigma_{\theta}^2$ and $\sigma_{\phi}^2$ are the corresponding error variances. }

Then, the transmit steering vector for {$i$}-UAV is given  by
\begin{equation}
\mathbf{a}(\theta^{\text{BS}}_i, \phi^{\text{BS}}_i)\in\mathbb{C}^{1\times T_x}= [
    1,\cdots,  e^{j\nu_T(T_x-1)\cos(\theta^{\text{BS}}_i)\cos(\phi^{\text{BS}}_i)}
],
\label{eq.steering}
\end{equation}
where $\nu_T=\frac{2\pi}{\lambda}d_T$ for $\lambda$ being  the waveform and $d_T$ being the spacing between adjacent antenna elements.
  Therefore, the received echo signal at the BS   from $i$-UAV target  over $L_s$ coherent time block becomes
\begin{equation}
    \mathbf{Y}_i = \beta_i \mathbf{a}(\theta^{\text{BS}}_i, \phi^{\text{BS}}_i)^\mathrm{H}\mathbf{a}(\theta^{\text{BS}}_i, \phi^{\text{BS}}_i)\mathbf{X}+\mathbf{N}_i,
    \label{eq.target_sensing}
\end{equation} where $\mathbf{X}=[\mathbf{x}(1), \cdots, \mathbf{x}(L_s)] $ and $\mathbf{N}_i\in\mathbb{C}^{T_x\times L_s}$ is the additive white Gaussian noise (AWGN) whose each entry follows $\mathcal{CN}(0,\sigma^2_{n})$ distribution. $\beta_i$ is the complex-valued round-trip path attenuation between the BS and $i$-UAV \cite{yu2023active}, which can be given as
\begin{equation}
\beta_i = \sqrt{\frac{\lambda^2 \Lambda}{64\pi^3 d_i^4}},
\end{equation}
 where $\Lambda$ is the radar cross section (RCS). Therefore, the sensing SINR of the $i$-UAV target can be calculated as
 \begin{equation}\label{eq:sensingSINR}
 \gamma_i = \frac{\left\|\beta_i \mathbf{a}(\theta^{\text{BS}}_i, \phi^{\text{BS}}_i)^\mathrm{H}\mathbf{a}(\theta^{\text{BS}}_i, \phi^{\text{BS}}_i)\mathbf{W} \right\|_{\mathrm{F}}^2}{\left\|\beta_j \mathbf{a}(\theta^{\text{BS}}_j, \phi^{\text{BS}}_j)^\mathrm{H}\mathbf{a}(\theta^{\text{BS}}_j, \phi^{\text{BS}}_j)\mathbf{W}\right\|_{\mathrm{F}}^2+\sigma_n^2},
 \end{equation}where $j\neq i\in\{L,E\}$.
 
{ For further sensing performance evaluation, the CRB, which serves as a fundamental lower bound on the variance of any unbiased estimator \cite{wang2023stars, jia2023physical}, is derived from the Fisher Information Matrix (FIM) for the AoD estimation of both the legitimate 
$L$-UAV and eavesdropper $E$-UAV targets.} In order to determine FIM of the $i$-UAV, first, the received echo signal given in (\ref{eq.target_sensing}) is vectorized as
\begin{align}
& \mathbf{y}_i = \mathbf{p}_i+\mathbf{v}_i,
\end{align}
 where $\mathbf{p}_i= \vect(\beta_i\mathbf{a}(\theta^{\text{BS}}_i, \phi^{\text{BS}}_i)^\mathrm{H}\mathbf{a}(\theta^{\text{BS}}_i, \phi^{\text{BS}}_i)\mathbf{X})$ and $\mathbf{v}_i=\vect(\mathbf{N}_i)$. Then, for $\boldsymbol{\omega}_i\in\mathbb{C}^{4\times 1}=[\boldsymbol{\vartheta}_i,\boldsymbol{\delta}_i]$, and $\boldsymbol{\vartheta}_i=[\theta^{\text{BS}}_i, \phi^{\text{BS}}_i]$ and $\boldsymbol{\delta}_i=[\mathfrak{Re} (\beta_i),  \mathfrak{Im}(\beta_i)]^\mathrm{T}$ the FIM for $i$-UAV  $\mathbf{F}_i\in\mathbb{C}^{4\times 4} $\cite{li2007range} can be expressed as follows
 \begin{align}
     \mathbf{F}_i=\begin{bmatrix}
         \mathbf{F}_{\boldsymbol{\vartheta}_i\boldsymbol{\vartheta}_i} & \mathbf{F}_{\boldsymbol{\vartheta}_i\boldsymbol{\delta}_i}\\
         \mathbf{F}_{\boldsymbol{\delta}_i\boldsymbol{\vartheta}_i} & \mathbf{F}_{\boldsymbol{\delta}_i\boldsymbol{\delta}_i}
     \end{bmatrix},
     \label{eq.fim}
 \end{align}
while the CRB estimation of AoD pairs of the $i$-UAV target is calculated as 
\begin{align}\label{eq:crb}
    & \text{CRB}(\boldsymbol{\vartheta}_i)=[\mathbf{F}_{\boldsymbol{\vartheta}_i\boldsymbol{\vartheta}_i}-\mathbf{F}_{\boldsymbol{\vartheta}_i\boldsymbol{\delta}_i}(\mathbf{F}_{\boldsymbol{\delta}_i\boldsymbol{\delta}_i})^{-1}\mathbf{F}_{\boldsymbol{\boldsymbol{\delta}_i\vartheta}_i}]^{-1}.
\end{align}
Each element of the FIM matrix in (\ref{eq.fim})   can be given as  
\begin{align}
&  \mathbf{F}_{i}(\epsilon, \tau)= \frac{2}{\sigma_n^2}\mathfrak{Re} \left(\frac{ {d} \mathbf{p}_i^\mathrm{H}}{d\boldsymbol{\omega}_{i,\epsilon}}\frac{ {d} \mathbf{p}_i}
{d\boldsymbol{\omega}_{i,\tau}}\right),
\end{align}
where  $\boldsymbol{\omega}_{i,\epsilon}$ and $\boldsymbol{\omega}_{i,\tau}$ are the $\epsilon$- and $\tau$-th elements of $\boldsymbol{\omega}_i$ for ${\epsilon, \tau}\in\{1,2,3,4\}$. 
{  It is worth noting that for the $E$-UAV target, since the horizontal and vertical angles  are   expressed as  $\phi^{\text{BS}}_E = \tilde{\phi}^{\text{BS}}_E + \Delta{\phi}_E$
 and $  \theta^{\text{BS}}_E = \tilde{\theta}^{\text{BS}}_E + \Delta{\theta}_E$, the steering vector $\mathbf{a}(\theta_E^{\text{BS}}\phi_E^{\text{BS}})$ in (\ref{eq.steering}) can be evaluated using the following trigonometric expansions:
 \begin{align}
 & \cos(\tilde{\theta}^{\text{BS}}_E + \Delta{\theta}_E)=\cos(\tilde{\theta}^{\text{BS}}_E)\cos( \Delta{\theta}_E)-\sin(\tilde{\theta}^{\text{BS}}_E)\sin( \Delta{\theta}_E)\nonumber,\\
& \cos(\tilde{\phi}^{\text{BS}}_E + \Delta{\phi}_E)=\cos(\tilde{\phi}^{\text{BS}}_E)\cos( \Delta{\phi}_E)-\sin(\tilde{\phi}^{\text{BS}}_E)\sin( \Delta{\phi}_E).
 \end{align}}

Therefore, for  $\mathbf{A}_i=\mathbf{a}^{\mathrm{H}}(\theta_i^{\text{BS}}\phi_i^{\text{BS}})\mathbf{a}(\theta_i^{\text{BS}}\phi_i^{\text{BS}})$, the FIM matrix elements become \cite{li2007range, Wang}
\begin{align}
    & \mathbf{F}_{\boldsymbol{\vartheta}_i\boldsymbol{\vartheta}_i}=\notag\\
    &\frac{2|\beta_i|^2L_s}{\sigma_n^2} \mathfrak{Re}\left(\begin{bmatrix}
    \Tr(\bar{\mathbf{A}}_{\theta_i^{\text{BS}}}\mathbf{R_x}(\bar{\mathbf{A}}_{\theta_i^{\text{BS}}} )^{\mathrm{H}})& \Tr(\bar{\mathbf{A}}_{\theta_i^{\text{BS}}}\mathbf{R_x}(\bar{\mathbf{A}}_{\phi_i^{\text{BS}}})^{\mathrm{H}})\\
    \Tr(\bar{\mathbf{A}}_{\phi_i^{\text{BS}}}\mathbf{R_x}(\bar{\mathbf{A}}_{\theta_i^{\text{BS}}})^{\mathrm{H}}) & \Tr(\bar{\mathbf{A}}_{\phi_i^{\text{BS}}}\mathbf{R_x}(\bar{\mathbf{A}}_{\phi_i^{\text{BS}}})^{\mathrm{H}})
    \end{bmatrix}\right),\\
&\mathbf{F}_{\boldsymbol{\delta}_i\boldsymbol{\vartheta}_i}=\frac{2L_s}{\sigma_n^2}\mathfrak{Re}\left(\begin{bmatrix}
{\beta_i^*\Tr(\mathbf{A}}_i\mathbf{R_x}(\bar{\mathbf{A}}_{\theta_i^{\text{BS}}} )^{\mathrm{H}})\\{\beta_i^*\Tr(\mathbf{A}}_i\mathbf{R_x}(\bar{\mathbf{A}}_{\phi_i^{\text{BS}}} )^{\mathrm{H}})
    \end{bmatrix}[
        1, j]\right),\\  &\mathbf{F}_{\boldsymbol{\delta}_i\boldsymbol{\delta}_i}=\frac{2L_s}{\sigma_n^2}\mathbf{I}_2\mathbf{R_x}
{\Tr(\mathbf{A}}_i\mathbf{R_x}(\bar{\mathbf{A}}_i )^{\mathrm{H}}),
\end{align}
where $\bar{\mathbf{A}}_{\theta_i^{\text{BS}}}$, $\bar{\mathbf{A}}_{\phi_i^{\text{BS}}}$ and $\bar{\mathbf{A}}_{\boldsymbol{\delta}_i}$ be the derivatives of  $\mathbf{A}_i$ with respect to $\theta_i^{\text{BS}}$, $\phi_i^{\text{BS}}$ and ${\boldsymbol{\delta}_i}$, respectively.  
 
\subsection{Communication  Model}
In the proposed scheme, the communication channels between the BS and RISs, as well as between the RISs and users, are modeled using Rician fading. Let $\mathbf{H}_n\in\mathbb{C}^{N_n\times T_x}$ denote the channel matrix between the BS and the $n$-th RIS, where $n\in\{L, M\}$ represents the legitimate ($L$-RIS) or malicious ($M$-RIS) surface. Similarly, let $\mathbf{g}_{n,k} \in \mathbb{C}^{1 \times N_n}$ denote the channel vector between the $n$-th RIS and single-antenna user $k$ (U$_k$) , for $k \in \{1, \cdots, K\}$. These channels can be expressed as 
\begin{align}
& \mathbf{H}_n = \sqrt{\frac{{L_0}}{d_{n}^{\alpha_{n}}}}\left({\sqrt{\frac{\kappa}{1+\kappa}}\mathbf{H}^{\text{LOS}}_n+\sqrt{\frac{1}{1+\kappa}}\mathbf{H}^{\text{NLOS}}_n}\right),\\
&\mathbf{g}_{n,k} = \sqrt{\frac{{L_0}}{d_{n,k}^{\alpha_{k}}}}\left(\sqrt{\frac{\kappa}{1+\kappa}}\mathbf{g}_{n,k}^{\text{LOS}}+\sqrt{\frac{1}{1+\kappa}}\mathbf{g}_{n,k}^{\text{NLOS}}\right),
\end{align}
where $\kappa$ denotes the Rician factor, $L_0$ is the reference path loss at a distance of $1$ meter (m), and $d_n$ and $d_{n,k}$ represent the distances between the BS and the $n$-RIS, and between the $n$-RIS and U$_k$, respectively. The corresponding path loss exponents are denoted by $\alpha_n$ and $\alpha_{n,k}$ for the BS–$n$-RIS and $n$-RIS–U$_k$ links, respectively. In the proposed scheme, each element of the NLOS components of the communication channels $\mathbf{H}^{\text{NLOS}}_n$ and  $\mathbf{g}^{\text{NLOS}}_{n,k}$ is assumed to be i.i.d. and following $\mathcal{CN}(0,1)$ distribution. On the other hand,  $\mathbf{H}^{\text{LOS}}_n$ and  $\mathbf{g}^{\text{LOS}}_{n,k}$ LOS components are deterministic and generated by steering vectors as follows:
\begin{align}
& \mathbf{H}^{\text{LOS}}_n = \mathbf{b}(\theta_n^\text{RIS},\phi_n^{\text{RIS}})^\mathrm{H}\mathbf{b}(\theta_n^\text{BS},\phi_n^{\text{BS}}),\\
&\mathbf{g}^{\text{LOS}}_{n,k} = \mathbf{b}(\theta_{n,k}^\text{RIS},\phi_{n,k}^{\text{RIS}}),
\label{eq.g_los}
\end{align}
where $\mathbf{b}(\theta_n^\text{BS},\phi_n^{\text{BS}})\in\mathbb{C}^{1\times T_x}$ is the ULA steering vector of the BS for the corresponding  $\{\theta_n^\text{BS},\phi_n^{\text{BS}}\}$ horizontal and vertical parts of the  2D AoD,  while  the UPA steering vectors of $n$-RIS towards BS and U$_k$ are  represented by $\mathbf{b}(\theta_{n}^\text{RIS},\phi_{n}^{\text{RIS}})\in\mathbb{C}^{1\times N_n}$ and $\mathbf{b}(\theta_{n,k}^\text{RIS},\phi_{n,k}^{\text{RIS}})\in\mathbb{C}^{1\times N_n}$, for the AoD of $\{\theta_{n}^\text{RIS},\phi_{n}^{\text{RIS}}\}$ and $\{\theta_{n,k}^\text{RIS},\phi_{n,k}^{\text{RIS}}\}$, respectively.  Therefore, they can be given as 
\begin{align}
& \mathbf{b}(\theta_n^\text{BS},\phi_n^{\text{BS}})=[1,\cdots,e^{j{\nu_T}(T_x-1)\cos(\theta^{\text{BS}}_n)\cos(\phi^{\text{BS}}_n)}],\\
&  \mathbf{b}(\theta_n^\text{RIS},\phi_n^{\text{RIS}})=\mathbf{b}^x(\theta_n^\text{RIS},\phi_n^{\text{RIS}})\otimes\mathbf{b}^z(\theta_n^\text{RIS},\phi_n^{\text{RIS}}),
\end{align}where $\mathbf{b}^x(\theta_n^\text{RIS},\phi_n^{\text{RIS}})\in\mathbb{C}^{1\times N_n^x}$ and $\mathbf{b}^z(\theta_n^\text{RIS},\phi_n^{\text{RIS}})\in\mathbb{C}^{1\times N_n^z}$ represent the steering vectors of the $n$-RIS along the $x$- and $z$-axes, respectively. $N_x^n$ and $N_z^n$ denote the number of reflecting elements along the $x$- and $z$-directions, and the total number of elements is given by $N_n = N_x^n \times N_z^n$ for $n\in\{L,M\}$. Therefore, $\mathbf{b}^x(\theta_n^\text{RIS},\phi_n^{\text{RIS}})$ and $\mathbf{b}^z(\theta_n^\text{RIS},\phi_n^{\text{RIS}})$
can be given as
\begin{subequations}
\begin{align}
&\mathbf{b}^x(\theta_n^\text{RIS},\phi_n^{\text{RIS}})=[1,\cdots,e^{j{\nu_R}(N_n^x-1)\cos(\theta^{\text{RIS}}_n)\cos(\phi^{\text{RIS}}_n)}],\\
&\mathbf{b}^z(\theta_n^\text{RIS},\phi_n^{\text{RIS}})=[1,\cdots,e^{j{\nu_R}(N_n^z-1)\sin(\phi^{\text{RIS}}_n)}], 
\end{align}  
\label{eq.steer-RIS}
\end{subequations}
\hspace{-12pt} where $\{\theta_n^\text{RIS},\phi_n^{\text{RIS}}\}$ is the AoD of the $n$-RIS towards BS and $\nu_R=\frac{2\pi}{\lambda}d_R$ for $d_R$ being the distance between two horizontally or vertically adjacent RIS elements. Similarly, the LOS component $\mathbf{g}_{n,k}^\text{LOS}(\theta_{n,k}^\text{RIS},\phi_{n,k}^{\text{RIS}})$ in (\ref{eq.g_los}) can be formulated as in (\ref{eq.steer-RIS}), based on the AoD pair $\{\theta_{n,k}^\text{RIS},\phi_{n,k}^{\text{RIS}}\}$.

The received signal at U$_k$, incorporating the reflections from both legitimate and malicious RISs, can be given as 
\begin{align}
&y_k(t)= \mathbf{g}_{L,k}\mathbf{\Omega}_L\mathbf{H}_L\mathbf{x}(t)+\mathbf{g}_{M,k}\mathbf{\Omega}_M\mathbf{H}_M\mathbf{x}(t)+n_k(t),
\label{eq.received}
\end{align}  
where $n_k$ is the AWGN figure with $\mathcal{CN}(0,\sigma_n^2)$ distribution. Therefore, the SINR of the U$_k$, given the overall beamforming matrix as $\mathbf{W}\in\mathbb{C}^{T_x\times (K+T_x)}=[\mathbf{w}_1, \cdots, \mathbf{w}_{K+T_x}]$, is expressed as
\begin{equation}  
\eta_k=\frac{\left|\mathbf{g}_{L,k}\mathbf{\Omega}_L\mathbf{H}_L\mathbf{w}_k\right|^2}{\sum_{\hat{k}}\left|\mathbf{g}_{L,k}\mathbf{\Omega}_L\mathbf{H}_L\mathbf{w}_{\hat{k}}\right|^2+\left\|\mathbf{g}_{M,k}\mathbf{\Omega}_M\mathbf{H}_M\mathbf{W}\right\|_\mathrm{F}^2+\sigma_n^2},  
\label{eq.com_SINR}
\end{equation}
where $\mathbf{w}_k\in\mathbb{C}^{T_x\times 1}$ is the corresponding beamforming vector towards U$_k$  for $k\in\{1,\cdots, K\}$ and $ {\hat{k}} \in\{1,\cdots, K+T_x\}$ such that ${\hat{k}}\neq k$.

\subsection{Security Model}
{ In the proposed scheme, the eavesdropper UAV target performs two separate functions: it (i) attempts to intercept the communication signals through a direct BS and UAV propagation channel, and (ii) carries a malicious RIS whose role is to degrade the reception at legitimate users through an interference attack. Accordingly, using (\ref{eq.steering}), the eavesdropping link is modeled as the direct link between the BS eavesdropper channel and the received signal at the eavesdropper UAV target at the $t$-th time index can be expressed as}
\begin{equation}
    y_E(t) = \sqrt{\beta_E}\mathbf{a}(\theta_E^{\text{BS}},\phi_E^{\text{BS}})\mathbf{x}(t)+n_E(t),
\end{equation}
where $n_E(t)$ is AWGN following $\mathcal{CN}(0,\sigma_n^2)$. Therefore, the eavesdropper SINR on the $k$-th legitimate user can be calculated as 
\begin{equation}  
    \eta_{E,k}=\frac{\left| \mathbf{a}(\theta_E^{\text{BS}},\phi_E^{\text{BS}})\mathbf{w}_k \right|^2}
   {\sum_{\hat{k}}\left| \mathbf{a}(\theta_E^{\text{BS}},\phi_E^{\text{BS}})\mathbf{w}_{\hat{k}} \right|^2+\sigma_n^2},
   \label{eq.eav_sinr}
\end{equation}
where $\hat{k}\in\{ 1,\cdots,K+T_x \}$ for $\hat{k}\neq k$. Hence, the secrecy rate of the $k$-th legitimate user can be expressed as 
\begin{equation} \label{eq:SR}
S_{R,k} = [R_k - R_{E,k}]^+ ,
\end{equation}
where $[x]^+=\max(x,0)$. $R_k=\log_2(1+\eta_k)$  is the achievable rate of user U$_k$ and $R_{E,k}=\log_2(1+\eta_{E,k})$ represents the eavesdropping rate at for U$_k$.

\section{Problem Formulation and Proposed Solution}

It is crucial to emphasize that in the proposed scheme, it is assumed that the BS  {  is unaware of the malicious RIS mounted on the $E$-UAV whose reflection matrix $\mathbf{\Omega}_M$ in (\ref{eq.m-ris}) using randomly generated phase shifts. Therefore, the
received signal at $U_k$ that perceived by the BS differs from (\ref{eq.received})  and can be given as \cite{alexandropoulos2023counteracting}
\begin{align}
&\hat{y}_k(t)= \mathbf{g}_{L,k}\mathbf{\Omega}_L\mathbf{H}_L\mathbf{x}(t)+n_k(t).
\label{eq.estimated}
\end{align} 
Therefore, using (\ref{eq.estimated}), the BS formulates the instantaneous  SINR of the U$_k$  as
\begin{equation} 
\hat{\eta}_k=\frac{\left|\mathbf{g}_{L,k}\mathbf{\Omega}_L\mathbf{H}_L\mathbf{w}_k\right|^2}{\sum_{\hat{k}}\left|\mathbf{g}_{L,k}\mathbf{\Omega}_L\mathbf{H}_L\mathbf{w}_{\hat{k}}\right|^2+\sigma_n^2}.
\label{eq.estiamed_sinr}
\end{equation}
Accordingly, using (\ref{eq.estiamed_sinr}), from the BS perspective, the secrecy rate of the $k$-th legitimate user is calculated as 
\begin{equation}
\hat{S}_{R,k} = [\hat{R}_k - R_{E,k}]^+
\label{eq.secrecy_rate_bs}
\end{equation}
 where $\hat{R}_k=\log_2(1+\hat{\eta}_k)$  is assumed to be the instantaneous achievable rate of U$_k$.

Therefore, based on the above calculations at the BS, the reflection coefficients of the legitimate RIS and the BS transmit beamforming matrix are jointly optimized to maximize the secrecy rate of the overall system. To address this, the following optimization problem is formulated:
}
\begin{subequations}
\begin{align}
& \text{(P1)}\hspace{0.4cm}\max_{\mathbf{W},\mathbf{\Omega}_L}\hspace{0.2cm} { \hat{S}_{R}}\\
&\hspace{1.5cm}\text{s.t.}\hspace{0.2cm}\Tr(\mathbf{WW}^\mathrm{H})\leq P_T,\\
&\hspace{2.3cm}\gamma_i\leq \rho_{s} \hspace{0.3cm}\text{for}\hspace{0.2cm}i\in\{L,E\},\\
& \hspace{2.3cm}|e^{j\varphi_l}|=1, \hspace{0.3cm}\text{for}\hspace{0.2cm}l\in\{1,\cdots,N_L\},
\end{align}
\label{eq.opt_joint}
\end{subequations}
\hspace{-10pt} where ${ \hat{S}_{R}=\sum\limits_{k=1}^{K}\hat{S}_{R,k}}$ is the achievable sum secrecy rate of legitimate users and  $\rho_{s}$ is the maximum achievable sensing SINR rate.  Since both legitimate and eavesdropper UAVs communicate directly with the BS without any reflected signals from RISs, the beamforming optimization can simply be independent of $L$-RIS configuration. This allows the problem (P1) to be decomposed into a two-stage optimization framework that first optimizing the transmit beamforming matrix $\mathbf{W}$, and later optimizing the $L$-RIS reflection matrix $\mathbf{\Omega}_L$. 

{  Accordingly, the joint design problem in (P1) involves the optimization of both the BS transmit beamforming matrix $\mathbf{W}$ and the $L$-RIS reflection matrix $\mathbf{\Omega}_L$. Because the direct BS-UAV links are not influenced by RIS reflections, the two variables become separable, enabling a sequential optimization framework. In the first stage, (P2) is formulated to optimize only the beamforming matrix $\mathbf{W}$ by maximizing the minimum sensing SINR of the two UAV targets under the transmit power and sensing quality constraints. Once the optimum beamforming matrix $\mathbf{W}$ is obtained, it yields a second-stage optimization denoted as (P3) that focuses solely on determining the constructive RIS matrix $\mathbf{\Omega}_L$ so as to maximize the achievable sum secrecy rate. This two-step decomposition cooperatively contributes to the overall goal of (P1).}

\subsection{Beamforming Optimization}
In the proposed scheme, in order to optimize the beamforming matrix $\mathbf{W}$ ensuring  achievable sensing performance for both legitimate and eavesdropper UAV targets, the following optimization problem is formulated:
\begin{subequations}
\begin{align} \label{eq.opt_beam}
    & \text{(P2)}\hspace{0.4cm} \max_{\mathbf{W}} { \min_i}\hspace{0.2cm}\gamma_i\\
    & \hspace{2cm}{\text{s.t.}}\hspace{0.2cm}\gamma_i\leq \rho_{s},\\ &\hspace{2.7cm}\Tr(\mathbf{WW}^\mathrm{H})\leq P_T.
\end{align}
  \label{eq.opt_beam1}
\end{subequations}
{  
In (P2), the sensing SINR of the $i$-UAV target is maximized while ensuring that the minimum sensing SINR remains above the acceptable threshold $\rho_{s}$, thereby guaranteeing reliable sensing performance.
}

For $\mathbf{A}_i\in\mathbb{C}^{T_x\times T_x}=\beta_i \mathbf{a}(\theta^{\text{BS}}_i, \phi^{\text{BS}}_i)^\mathrm{H}\mathbf{a}(\theta^{\text{BS}}_i, \phi^{\text{BS}}_i)$ being the round trip channel between BS and $i$-UAV target,    then sensing SINR $\gamma_i$ in (\ref{eq:sensingSINR}) can be re-expressed in the following quadratic form  
\begin{align}
    & \gamma_i = \frac{\Tr(\mathbf{A}_i^\mathrm{H}\mathbf{A}_i\mathbf{R_x})}{\Tr(\mathbf{A}_j^\mathrm{H}\mathbf{A}_j\mathbf{R_x})+\sigma_{n}^2},
\end{align}
where $j\neq i\in \{L, E\}$. Therefore, using a SDR-based approach, 
(P2) can be reformulated as
\begin{subequations}
    \begin{align}
        & \text{(P2)}\hspace{0.4cm} \max_\mathbf{R_x}\hspace{0.1cm}{ \min_i}\hspace{0.2cm}\gamma_i\\
   & \hspace{1.9cm}{\text{s.t.}}\hspace{0.3cm}\gamma_i\leq \rho_{s},\\
&\hspace{2.6cm}\Tr(\mathbf{A}_L^\mathrm{H}\mathbf{A}_L\mathbf{R_x})\geq\Tr(\mathbf{A}_E^\mathrm{H}\mathbf{A}_E\mathbf{R_x})+\sigma_n^2,\\
&\hspace{2.6cm}\Tr(\mathbf{R_x})\leq P_T.
    \end{align}
    \label{eq.beam_opt2}
\end{subequations}
{ It is important to note that, as reflected in constraint (\ref{eq.beam_opt2}c), the BS prioritizes the 
$L$-UAV target while maintaining a minimum sensing SINR threshold $\rho_s$
for the $E$-UAV.   After that, using the SeDuMi solver within the CVX optimization toolbox \cite{cvx}, (P2) can be effectively solved. Subsequently,  eigenvalue decomposition (EVD) is applied to recover the beamforming matrix $\mathbf{W}$
\cite{zhang2017matrix}.

Moreover, in order to evaluate the spatial selectivity of the optimized beamforming matrix, ensuring that the transmitted energy is effectively concentrated toward the desired UAV target, $L$-UAV, while minimizing interference in other directions, the beampattern gain of the $i$-th target in direction of $\{\theta^{\text{BS}}_i, \phi^{\text{BS}}_i\}$ angles can be calculated using steering vectors $\mathbf{a}(\theta^{\text{BS}}_i, \phi^{\text{BS}}_i)$  in (\ref{eq.target_sensing}) as
\begin{equation}
\mathcal{P}_i= \mathbf{a}(\theta^{\text{BS}}_i, \phi^{\text{BS}}_i)^\mathrm{H}\mathbf{R_x}\mathbf{a}(\theta^{\text{BS}}_i, \phi^{\text{BS}}_i). 
\end{equation}

}
\subsection{Legitimate RIS Phase Optimization}
In this subsection, after the beamforming matrix $\mathbf{W}$ is obtained, the BS optimizes the reflection coefficients of $L$-RIS  to maximize the secrecy rate of the $K$ legitimate users.  {  At the BS, the secrecy rate $\hat{S}_R$ in (\ref{eq.secrecy_rate_bs}) is computed as the total sum of achievable secrecy rates across all legitimate users. Since}  eavesdropper SINR of $E$-UAV target (\ref{eq.eav_sinr}) across users are independent from $L$-RIS reflection, the $\boldsymbol{\Omega}_L$ (\ref{eq.l-ris}) can be optimized to enhance { $\hat{S}_R$} by maximizing sum of achievable rates of all users {  $\sum_k \hat{R}_k$}. Therefore, { considering (\ref{eq.estiamed_sinr})}, the optimization of $\mathbf{\Omega}_L$  in (P1) can be converted to following problem:
\begin{subequations}
    \begin{align}
& \text{(P3)}\hspace{0.4cm}  \max_{\mathbf{\Omega}_L }\hspace{0.1cm}{ \min_k}\hspace{0.1cm} { \hat{\eta}_k}\\
& \hspace{1.3cm}\text{s.t.}\hspace{0.2cm}
|e^{j\varphi_l}|=1\hspace{0.2cm}\text{for}\hspace{0.2cm}l\in\{1,\cdots,N_L\}\label{eq.unitmodu}.    \end{align}
\label{eq.opt_phase1}
\end{subequations}
\hspace{-8pt} However, due to its non-concave objective function and non-convex unit-modulus constraints,  the problem (P3) is difficult to solve \cite{zhou2020intelligent}. Therefore, an SDR-based approach is adopted to simplify (\ref{eq.opt_phase1}). 

As a first step, { the instantaneous communication SINR given in (\ref{eq.estiamed_sinr}) }is re-expressed in a quadratic form by applying appropriate mathematical modifications. Applying trace equality \cite{zhang2017matrix}, the numerator in { (\ref{eq.estiamed_sinr}) }
can be rewritten as $\left|\mathbf{g}_{L,k}\mathbf{\Omega}_L\mathbf{H}_L\mathbf{w}_k\right|^2=\Tr(\mathbf{\bar{H}}_{L,k}\mathbf{\Omega}_L^\mathrm{H}\mathbf{\bar{G}}_{L,k}^\mathrm{T}\mathbf{\Omega}_L)=\mathbf{z}_L(\mathbf{\bar{H}}_{L,k}\odot\mathbf{\bar{G}}_{L,k})\mathbf{z}_L^\mathrm{H}$, where $\mathbf{\bar{H}}_{L,k}\in\mathbb{C}^{N_L\times N_L}=\mathbf{H}_L\mathbf{w}_k\mathbf{w}_k^\mathrm{H}\mathbf{H}_L^\mathrm{H}$ and $\mathbf{\bar{G}}_{L,k}^\mathrm{T}\in\mathbb{C}^{N_L\times N_L}=\mathbf{g}_{L,k}^\mathrm{H}\mathbf{g}_{L,k}$, while $\mathbf{z}_L\in\mathbb{C}^{1\times N_L}$ is the reflection vector of $L$-RIS that composes non-zero diagonal elements of reflection matrix $\mathbf{\Omega}_L$. In a similar way,  the denominator of 
{ $\hat{\eta}_k$} can be rewritten in quadratic form as  $\left|\mathbf{g}_{L,k}\mathbf{\Omega}_L\mathbf{H}_L\mathbf{w}_{\hat{k}}\right|^2=\mathbf{z}_L(\mathbf{\bar{H}}_{L,k}\odot\mathbf{\bar{G}}_{L,\hat{k}})\mathbf{z}_L^\mathrm{H}$ 
where $\mathbf{\bar{G}}_{L,\hat{k}}^\mathrm{T}\in\mathbb{C}^{N_L\times N_L}=\mathbf{g}_{L,\hat{k}}^\mathrm{H}\mathbf{g}_{L,\hat{k}}$, 
Therefore, { (\ref{eq.estiamed_sinr})} can be rewritten as
{ \begin{align}
 &  \hat{\eta}_k =\frac{\Tr(\mathbf{C}_{L,k}\mathbf{Z}_L)}{\sum_{\hat{k}}\Tr(\mathbf{C}_{L,\hat{k}}\mathbf{Z}_L)+\sigma_n^2},   
\end{align}}\hspace{-3pt}
where $\mathbf{C}_{L,k}\in\mathbb{C}^{N_L\times N_L}=\mathbf{\bar{H}}_{L,k}\odot\mathbf{\bar{G}}_{L,k}$, $\mathbf{C}_{L,\hat{k}}=\mathbf{\bar{H}}_{L,k}\odot\mathbf{\bar{G}}_{L,\hat{k}}$, 
while $\mathbf{Z}_L\in\mathbb{C}^{N_L\times N_L} = \mathbf{z}_L^{\mathrm{H}}\mathbf{z}_L $ 
Therefore, the legitimate phase optimization problem in (\ref{eq.opt_phase1}) can be converted to the following quadratically constrained quadratic programming (QCQP) problem as follows
{ 
\begin{subequations}
\begin{align}
& \text{(P3)}\hspace{0.3cm}\max_{\mathbf{Z}_L}\hspace{0.2cm}{ \min_k}\hspace{0.1cm} \hat{\eta}_k \\
& \hspace{1.1cm}\text{s.t.} \hspace{0.2cm} \Tr(\mathbf{C}_{L,k}\mathbf{Z}_L) - \Big( 
    \sum_{\hat{k}} \Tr(\mathbf{C}_{L,\hat{k}}\mathbf{Z}_L) + \sigma_n^2
  \Big) \geq 0, \\
&  \hspace{1.8cm}\mathbf{Z}_L \succeq 0, \\
& \hspace{1.8cm}\mathbf{Z}_L(l,l)=1\hspace{0.2cm}\text{for}\hspace{0.2cm}l\in\{1,\cdots,N_L\}\label{eq.sdr}.
\end{align}
\label{eq.opt_phase2}
\end{subequations}}
\hspace{-8pt} After relaxing the non-convex constraint   (\ref{eq.sdr}) via an SDR-based approach, (P3) can be efficiently solved using the CVX optimization toolbox \cite{cvx}. However, since the obtained solution is not always guaranteed to be rank-one, additional Gaussian approximation or EVD can be applied to get a feasible rank-one solution \cite{luo2010semidefinite}.
{  Overall, after transmit beamforming matrix and legitimate RIS phase shifts are respectively obtained from  (\ref{eq.beam_opt2}) and (\ref{eq.opt_phase2}), the exact achievable secrecy rate of the legitimate system via (\ref{eq:SR}), leveraging the communication and eavesdropper SINR expressions in (\ref{eq.com_SINR}-\ref{eq.eav_sinr}).}

{  
\subsection{Malicious RIS Optimization for Worst-Case Scenario}
In this section, to evaluate the robustness of the proposed dual target mounted RISs-assisted secure ISAC framework, a worst-case malicious interference scenario is examined in which the hostile system controlling malicious RIS is assumed to have perfect knowledge of the beamforming matrix $\mathbf{W}$, the perfect CSI between the BS–$M$-RIS channel $\mathbf{H}_M$ and the $M$-RIS–U$_k$ channel $\mathbf{g}_{M,k}$.
Under this worst-case model, the malicious RIS optimizes its phase shift vector to intentionally degrade the received signal power at the legitimate users. Specifically, given the acquired CSI and the legitimate RIS configuration, a dedicated second-stage optimization to minimize the effective channel gain of each legitimate user. This analysis provides insight into the performance limits of the proposed system when subjected to an optimally coordinated interference attack. Then, to disrupt the legitimate communication, the following optimization problem is formulated:
\begin{subequations}
\begin{align}
& \text{(P4)}\hspace{0.3cm} \max_{\mathbf{\Omega}_M}\hspace{0.2cm} \left\|\mathbf{g}_{M,k}\mathbf{\Omega}_M\mathbf{H}_M\mathbf{W}\right\|_\mathrm{F}^2  \\ &\hspace{1cm}\text{s.t.}\hspace{0.2cm}
|e^{j\zeta_m}|=1, \hspace{0.3cm}\text{for}\hspace{0.2cm}m\in\{1,\cdots,N_M\}.
\end{align}\label{eq:worst}
\end{subequations}

Following the optimization of the worst-case malicious RIS interference in (\ref{eq:worst}), its impact on the communication SINR in (\ref{eq.com_SINR}) is incorporated, and the resulting secrecy rate of the legitimate system under these optimized malicious attacks is evaluated using (\ref{eq:SR}).

\subsection{Computational Complexity}
The computational complexity of the proposed scheme is primarily determined by the SDR-based solution employed in the two-stage optimization process. The first stage's computational complexity of the SDR-based solution for the beamforming optimization problem (P2) in (\ref{eq.beam_opt2}) depends on the number of transmit antennas and solution accuracy. For $\varepsilon>0$ being the solution accuracy, the complexity is calculated as $\mathcal{O}(T_x^{4.5}\log(1/\varepsilon)$. The second stage's computational complexity of the SDR-based solution for the legitimate RIS phase optimization (P3) in (\ref{eq.opt_phase2}) depends on the number of RIS elements and solution accuracy. Namely, this stage entails a complexity of $\mathcal{O}(N_L^{4.5}\log(1/\varepsilon))$ \cite{luo2010semidefinite}.
}





\begin{figure}[t]
    \centering
    \includegraphics[width=1\linewidth]{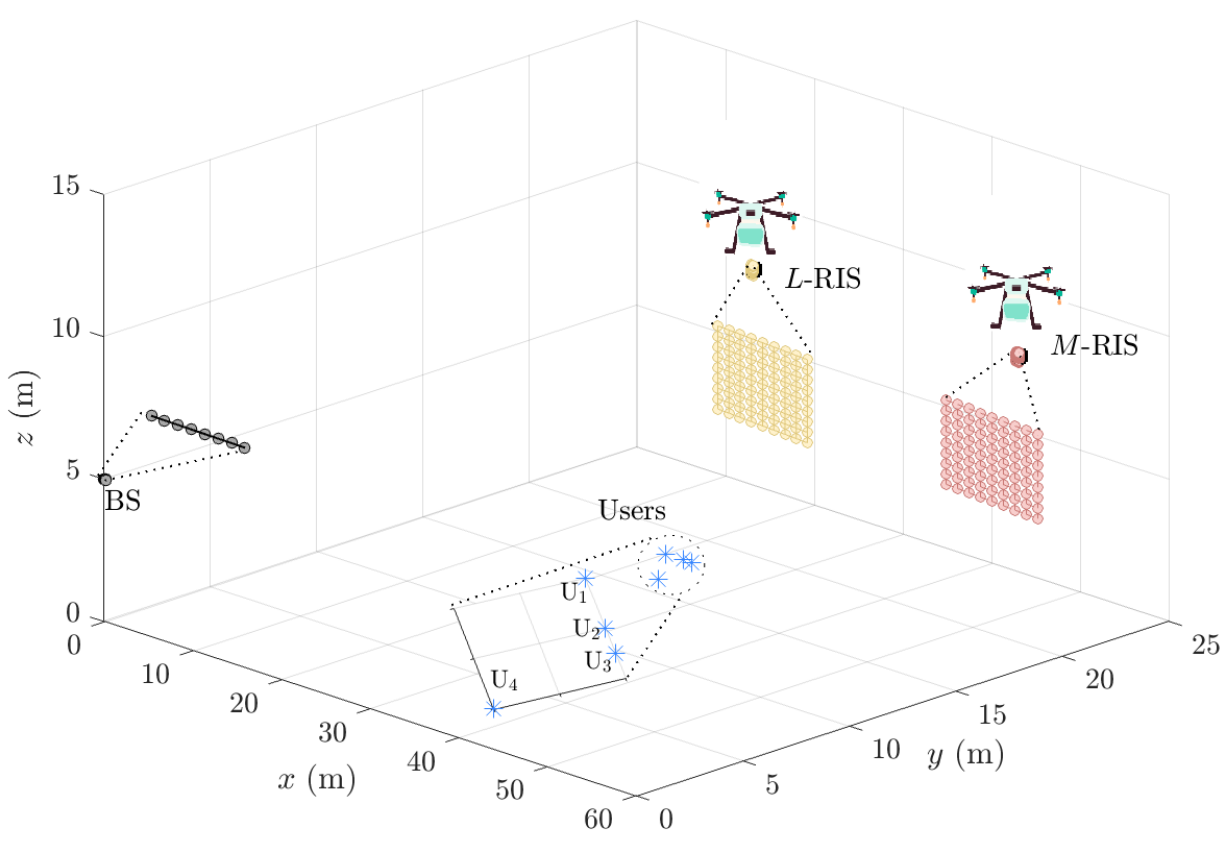}
    \caption{Spatial layout of the simulation environment for the BS, the RISs, and the users.}
    \label{Fig:locs}
\end{figure}
 \begin{table}[t]
\centering
\renewcommand{\arraystretch}{1.5}
\caption{Simulation Parameters}
\label{Tab:params}
\resizebox{\linewidth}{!}{
\begin{tabular}{|c|c|l|}
\hline
\textbf{Parameter} & \textbf{Value} & \textbf{Description} \\
\hline\hline
$f_c$ & $3.5$ GHz & Carrier frequency \\\hline $T_x$ & $8$ & Number of transmit antennas \\ \hline
$K$ & $4$ & Number of legitimate users \\\hline
$M$ & $2$ & Number of UAV sensing targets \\\hline
$L_s$ & $100$ & Coherent time length\\\hline
$\kappa$ & $4$ dB & Rician factor\\\hline
$\alpha_n$, $\alpha_k$ & $2.2$ & Path loss exponent\\\hline
$S$ & $1$ $\text{m}^2$ & RCS \\\hline
$L_0$ & $-30$ dB & Reference path attenuation at $1$ m\\\hline
$ \sigma^2_n$ & $-120$ dBW & Noise power \\\hline
$\theta^{\text{BS}}_L,\phi^{\text{BS}}_L$ &  $(-141^\circ,-9^\circ)$ &  AoD pairs for $L$-UAV target\\\hline
${ \tilde{\theta}^{\text{BS}}_E,\tilde{\phi}^{\text{BS}}_E}$ &${ (-160^\circ,-5^\circ)}$  &{   AoD pairs for $E$-UAV target}\\\hline
${ \sigma_{\theta}^2, \sigma_{\phi}^2}$ &${ (5^\circ)^2,(5^\circ)^2}$  &  { Error variances} \\\hline
$d_L$ &  $32.4$ m &  Distance from BS to $L$-UAV \\\hline
$d_E$ &  $58.7$ m &  Distance from BS to $E$-UAV \\\hline
${d_{L,k}}$ & $\{11.3, 10.6, 10.4,   10.8\} $ m& Distance from $L$-RIS to  ${\text{U}_k}$, for $k\in\{1:K\}$ \\\hline
$d_{M,k}$ &  $\{36.4, 34.5, 33.6, 32.8 \}$ m &  Distance from $M$-RIS to U$_k$, for $k\in\{1:K\}$\\\hline
\end{tabular}
}
\end{table}
\section{Performance Evaluation}
In this section, the secrecy rate and sensing performance of the proposed UAV-mounted RIS-assisted secure ISAC system are evaluated through Monte Carlo simulations under various scenarios. A BS with $T_x=8$ transmit antennas, $K=4$ legitimate users and two UAV sensing targets, where one is equipped with a legitimate RIS and the other with a malicious RIS, are considered. The BS is positioned at a height of $5$ m above ground level, and both UAV targets are deployed at $10$ m height. The users are randomly distributed at ground level, where the details of the placements are illustrated in Fig. \ref{Fig:locs}. Additionally, the simulation parameters are summarized in Table \ref{Tab:params}. 


\begin{figure}[t]
    \centering
\includegraphics[width=1\linewidth]{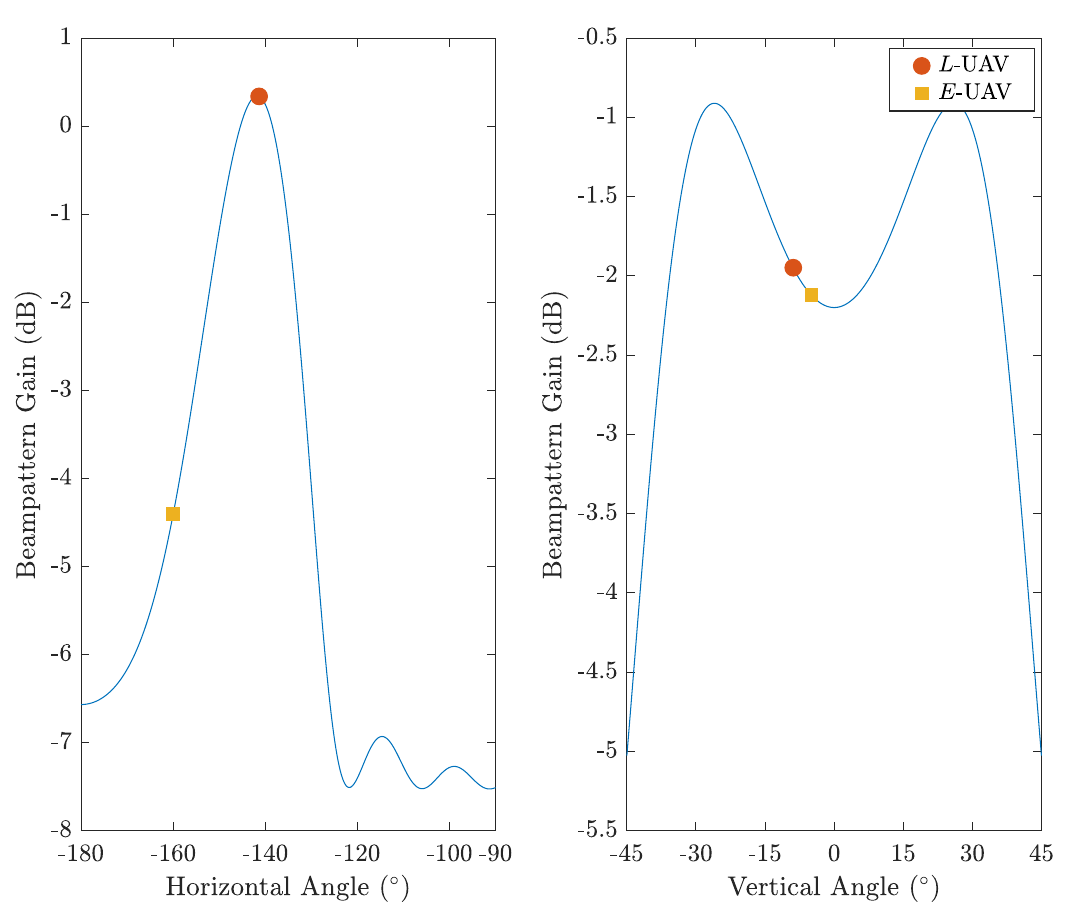}
    \caption{{ Beampattern gain of the proposed dual-target RISs-assisted ISAC scheme.}}
    \label{Fig:beampattern}
\end{figure}
\begin{figure}[t]
    \centering    \includegraphics[width=1\linewidth]{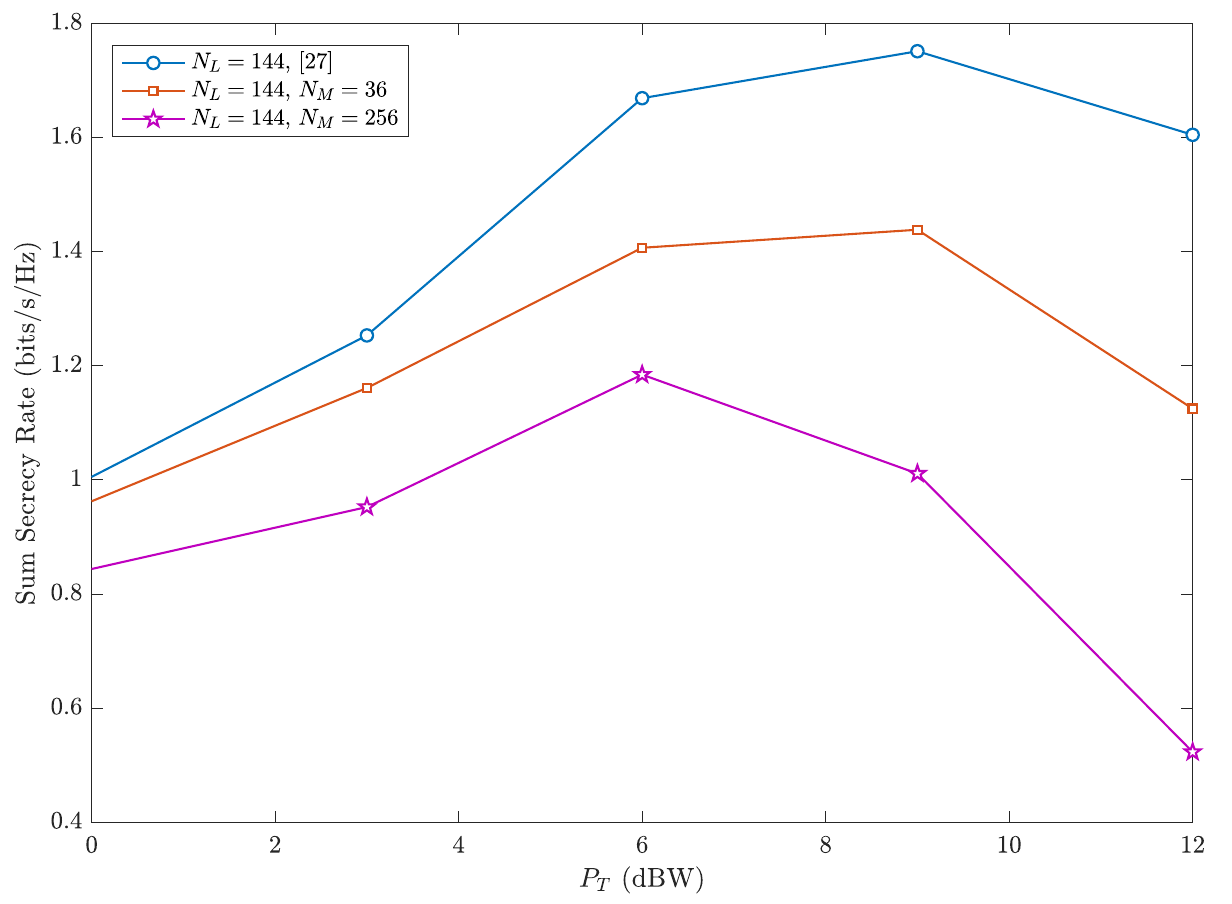}
    \caption{{ Comparison of the sum secrecy rate of the proposed scheme with reference \cite{benchmark} for varying $N_M$ sizes.}}
    \label{Fig:MRIS}
\end{figure}

{  
Fig. \ref{Fig:beampattern} illustrates the beampattern gain of the proposed scheme across both horizontal and vertical angles at $P_T=-4$ dBW. Since the beamforming matrix is optimized in  (\ref{eq.beam_opt2}) to maximize the minimum sensing SINR of the UAV targets, directing the beams primarily toward the $L$-UAV target, it is clearly seen from the results that the beampattern exhibits a distinct peak at the $\theta_L^\mathrm{BS}$ horizontal angle of $L$-UAV target. Conversely, the gain across the vertical angles remains relatively similar for both targets, as their elevation coordinates are nearly identical.

Fig. \ref{Fig:MRIS} illustrates the comparison of the $S_R$	performance between the proposed  scheme and the RIS-aided ISAC system in \cite{benchmark}, which serves as the benchmark. Both schemes adopt the same system configuration with two targets while one target behaves as an eavesdropper. However,   no malicious RIS is assumed in benchmark scheme  \cite{benchmark}. The results show that the proposed scheme, operating under dual security threats from an eavesdropper target and a malicious RIS attacks, and the benchmark scheme in \cite{benchmark} under a eavesdropper target threat, exhibit similar $S_R$	trends, where the benchmark scheme demonstrates a better $S_R$ performance. Moreover, it is observed that when $P_T$ increases beyond a certain point, the $S_R$ performance starts degrading. These results can be attributed to the fact that, although a higher $P_T$ improves the achievable rate of the legitimate users, it simultaneously increases the sensing interference experienced by the users and enhances the SINR of the $E$-UAV target. This highlights the critical trade-off between enhancing legitimate user rates and mitigating interference effects, which necessitates the appropriate selection of $P_T$ values. Furthermore, it can be deduced from the results that the proposed scheme  sustains reliable communication under increasing malicious attacks.}


\begin{figure}[t]
    \centering
\includegraphics[width=1\linewidth]{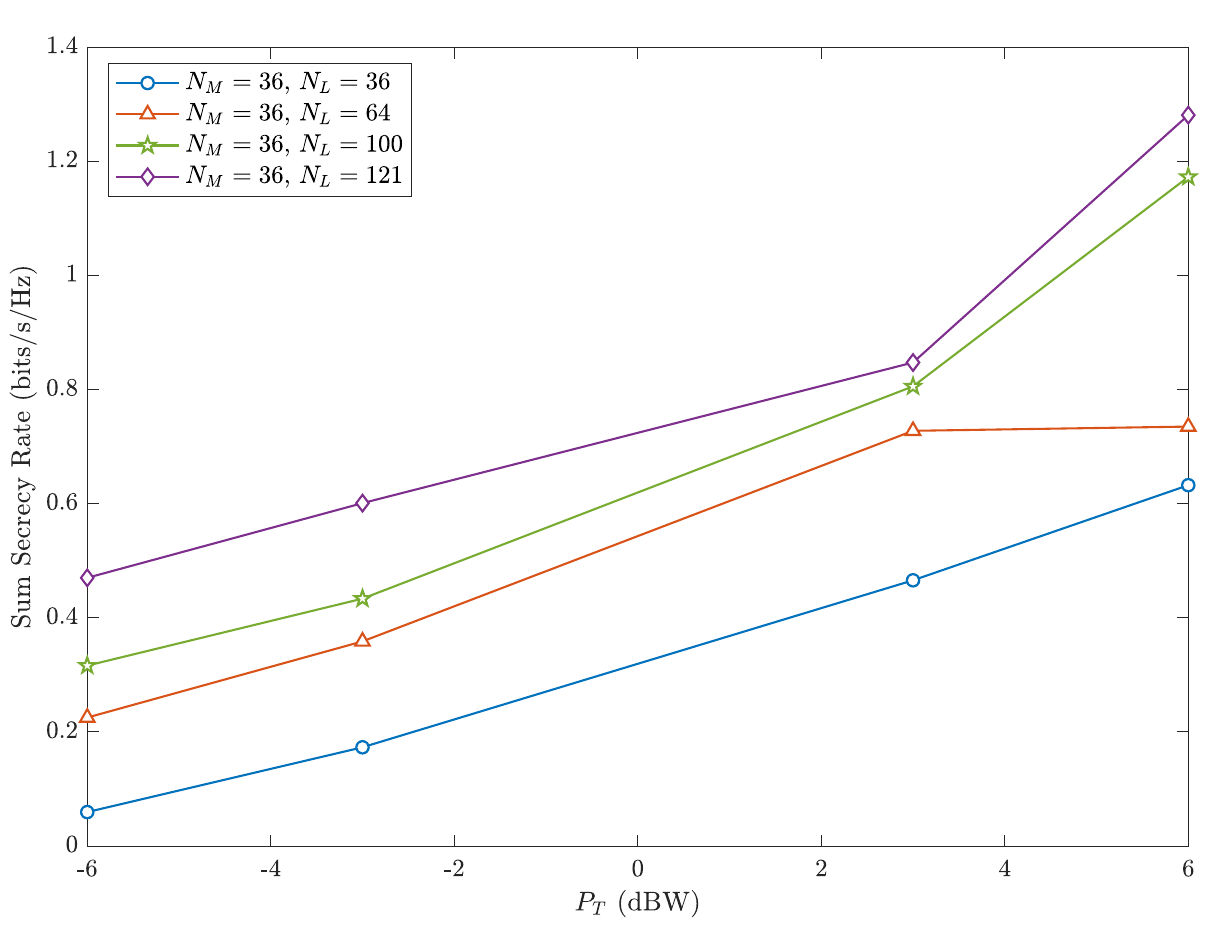}
\caption{{ Comparison of sum secrecy rate of the proposed scheme for varying $N_L$ sizes.}}
    \label{Fig:LRIS}
\end{figure}

\begin{figure}[t]
    \centering    
\includegraphics[width=1\linewidth]{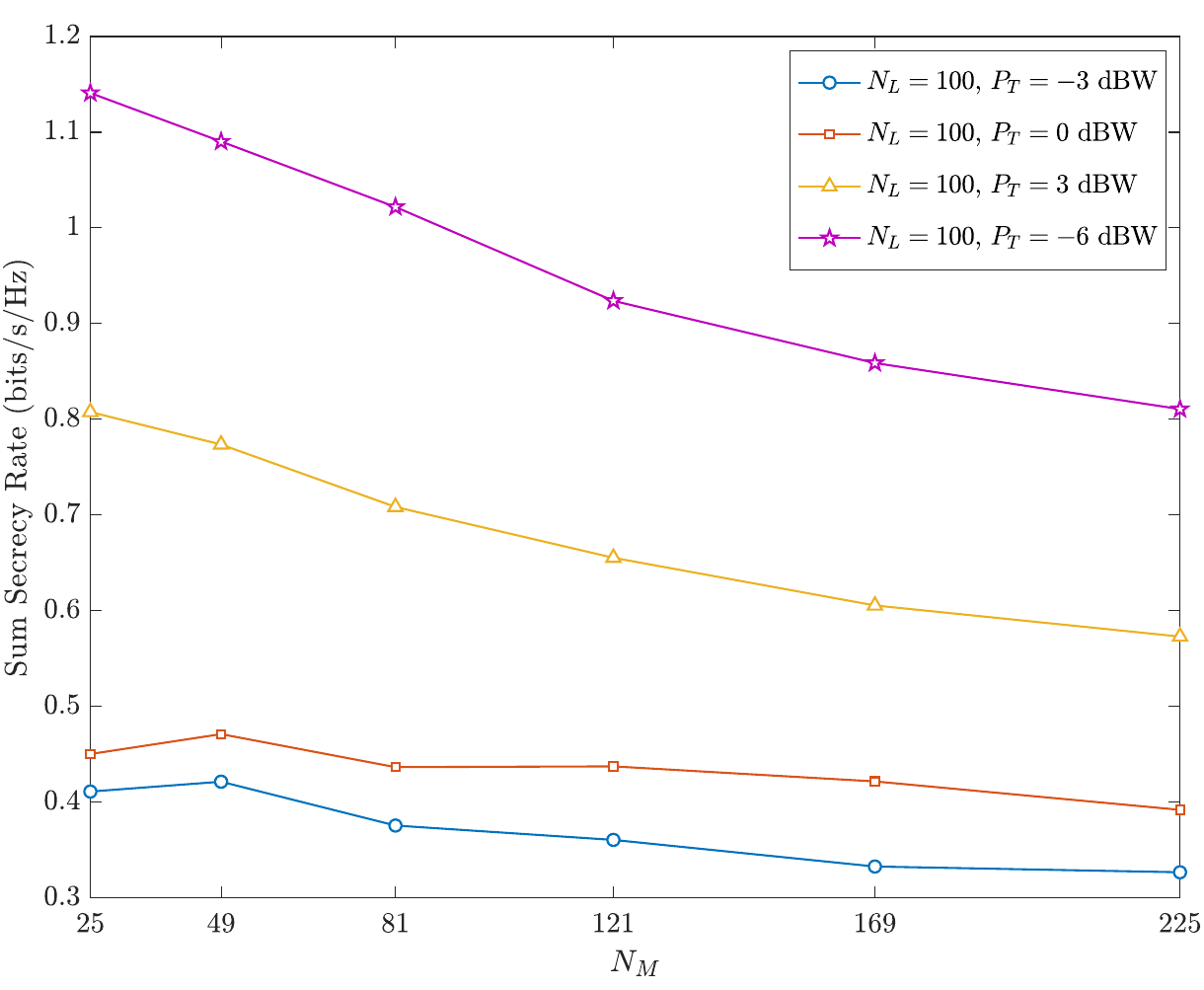}
    \caption{{ Sum secrecy rate performance of the proposed scheme for increasing $N_M$ sizes.}}
    \label{Fig:num_MRIS}
\end{figure}
\begin{figure}[t]
    \centering   \includegraphics[width=1\linewidth]{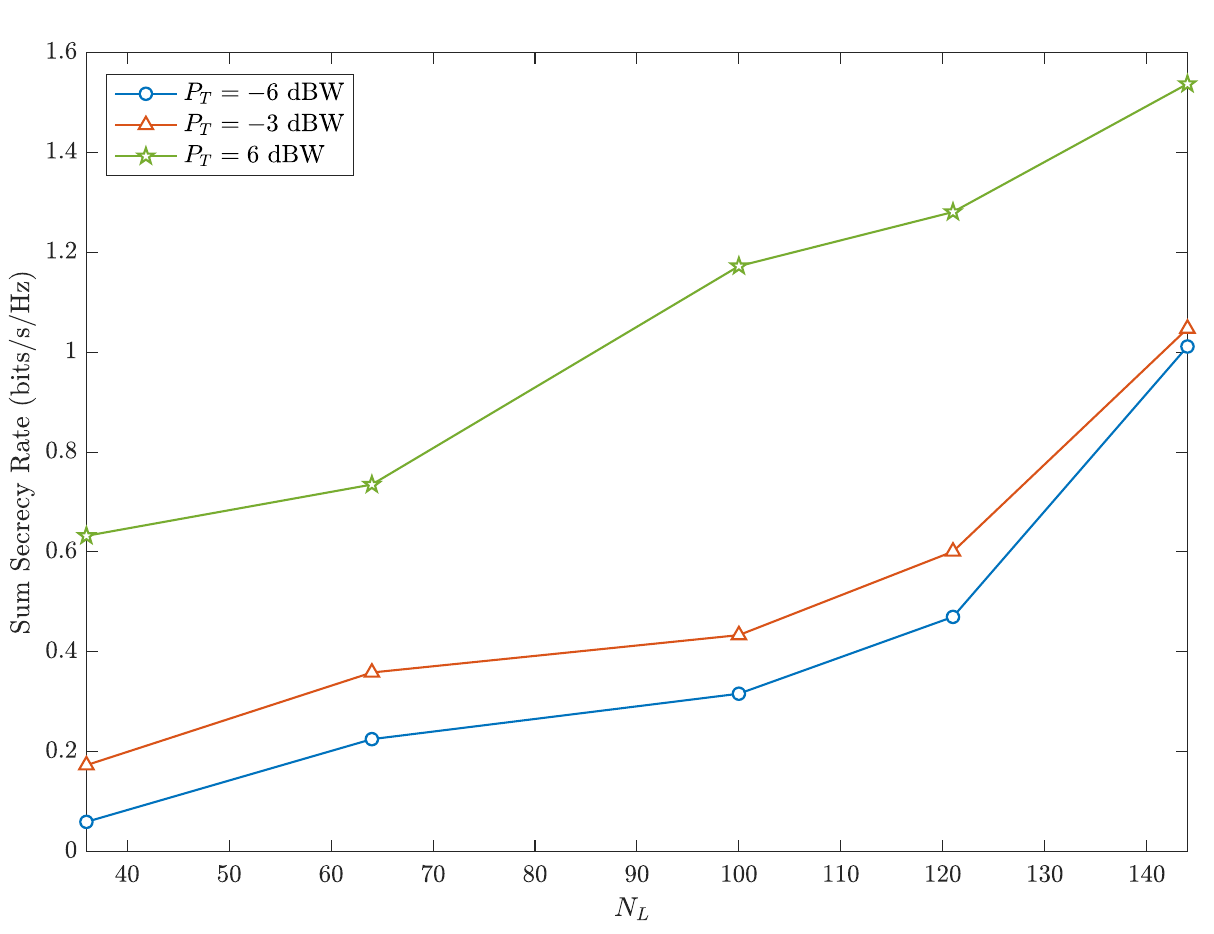}
    \caption{{ Sum secrecy rate performance of the proposed scheme for varying $P_T$ values.}}
    \label{Fig:num_LRIS}\vspace{-10pt}
\end{figure}

{ 
Fig. \ref{Fig:LRIS} illustrates the $S_R$ performance of the proposed scheme with a fixed malicious RIS size of $N_M=36$, which is evaluated for varying $N_L$. As expected, the secrecy rate generally improves with higher $P_T$, which provides a larger power budget to simultaneously meet sensing constraints and overcome interference. Moreover, increasing $N_L$ yields substantial performance gains. A larger legitimate RIS provides greater spatial degrees of freedom and higher passive beamforming resolution.}


{  In Fig. \ref{Fig:num_MRIS}, $S_R$ performance of the proposed system for $N_L=100$  and different  $N_M\in\{25\sim225\}$ is presented. Although the improvement is modest at low $P_T$, the results show that   $S_R$
 of the proposed system improves as $P_T$
 increases. Moreover, it can be observed that larger values of $N_M$ lead to a degradation in the secrecy rate performance.  The results also reveal that the proposed scheme maintains secure communication even at relatively low $P_T$
 values and large $N_M$ cases, thereby demonstrating the robustness of the proposed algorithm against both eavesdropping and malicious attacks.}


\begin{figure}[t]
    \centering
    \vspace{-0.4cm}
    \subfloat[]{%
      \includegraphics[width=0.5\linewidth, height=0.8\linewidth]{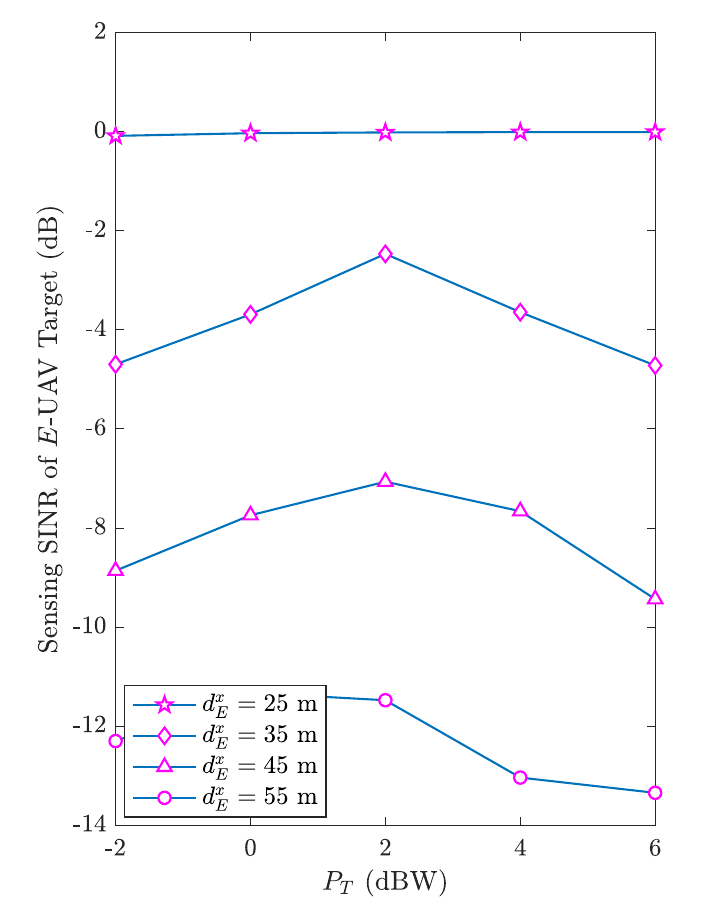}
    \label{fig.sinr.a}}
    \subfloat[]{%
   \includegraphics[width=0.5\linewidth, height=0.81\linewidth]{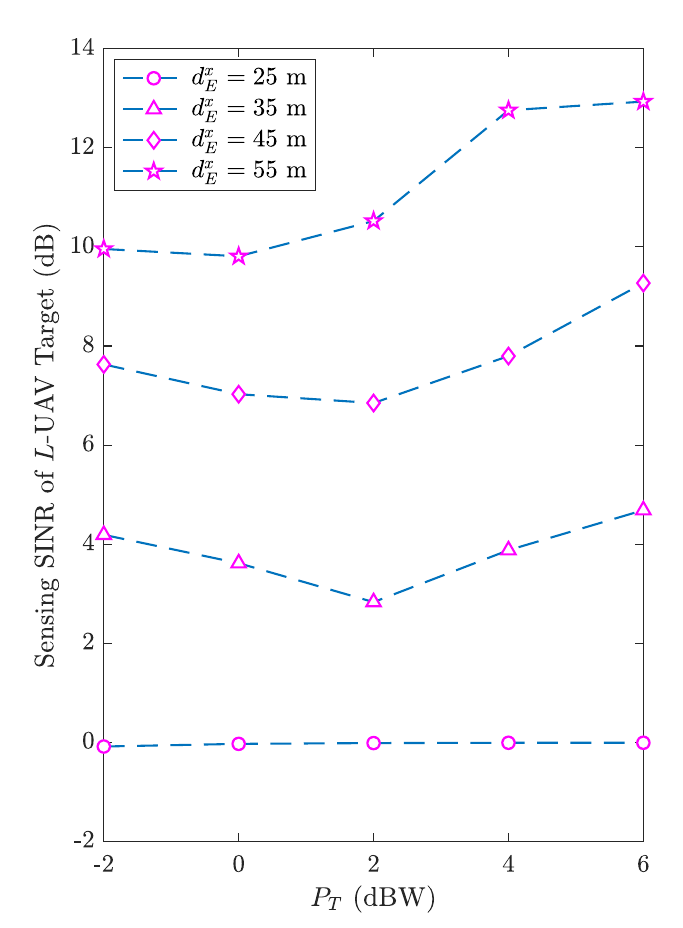}
    \label{fig.sinr.b}}
    \caption{{ Sensing SINR of both UAV targets for varying horizontal distance $d^x_E$.}}
    \label{fig.sinr}\vspace{-10pt}
\end{figure}

{ 
Fig. \ref{Fig:num_LRIS} illustrates the $S_R$ performance of the proposed scheme for increasing $N_L$ sizes. As observed, the secrecy rate increases with $N_L$ for all considered power levels. This highlights that scaling the legitimate RIS provides spatial degrees of freedom, translating to a higher passive beamforming gain. Consequently, the legitimate system can more effectively direct signals to intended users, even when the transmit power is strictly limited (e.g., at $P_T = -6$ dBW). Furthermore, an increase in $P_T$ improves the system's performance across all $N_L$ configurations, as the BS gains a larger power margin to simultaneously satisfy the dual objectives of secure communication and high-resolution sensing.} 

\begin{figure}[t]
    \centering
    \includegraphics[width=1\linewidth]{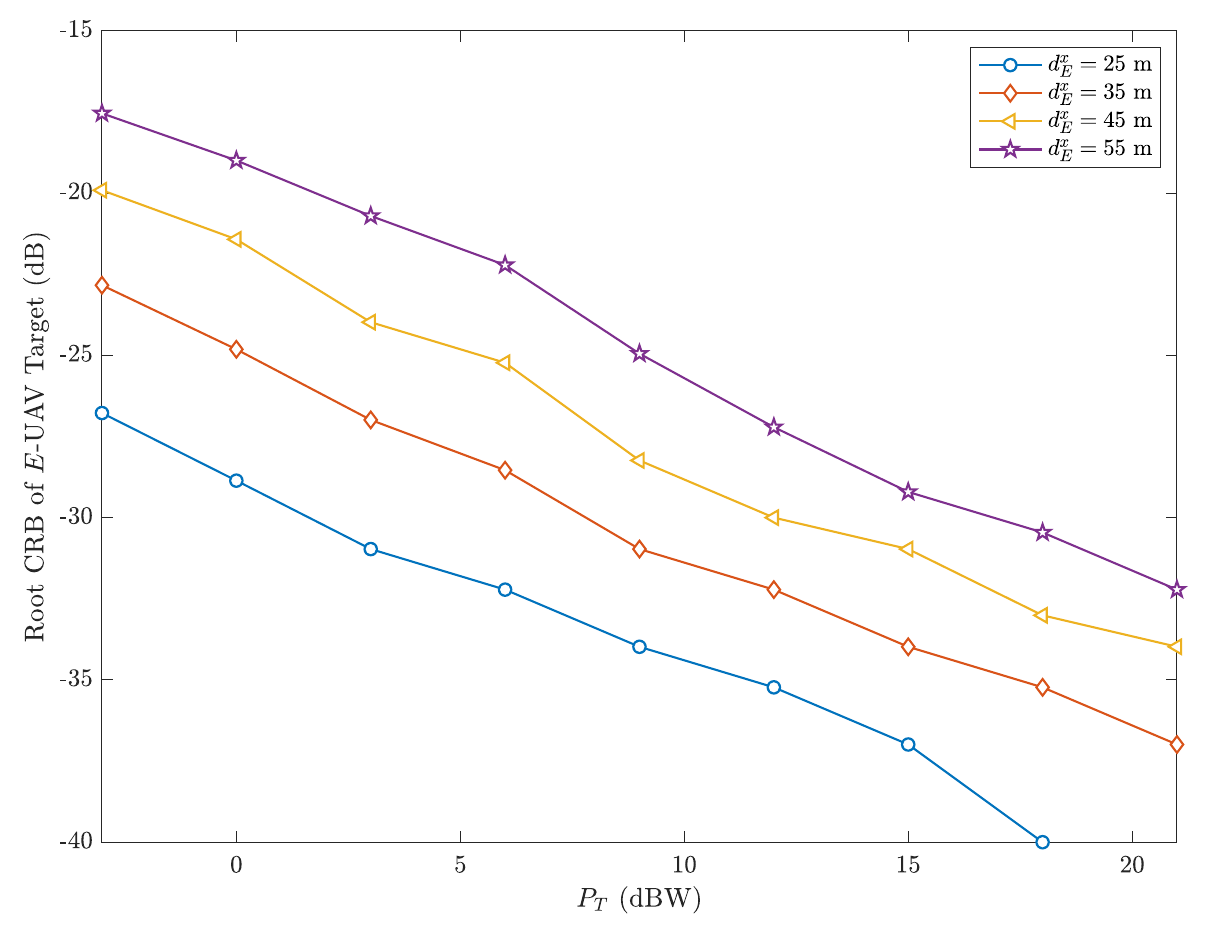}
    \caption{{ Root CRB of the $E$-UAV for varying horizontal distance $d^x_E$.}}
    \label{Fig:CRB} \vspace{-10pt}
\end{figure}

{ In the following, the sensing performance of the proposed system is evaluated in terms of sensing SINR of both the legitimate and the eavesdropper targets, and the root CRB estimations for AoDs of the eavesdropper UAV target, as illustrated in Figs. \ref{fig.sinr} and \ref{Fig:CRB}, respectively.  It should be emphasized that the CRB is utilized in the results as a conclusive evaluation benchmark to validate the effectiveness of the SINR-based beamforming design, rather than serving as an active constraint during the optimization phase.} 

{ In Fig. \ref{fig.sinr}, to evaluate the effectiveness of the proposed algorithm on the sensing SINR performance of UAV targets, the SINR values for both the legitimate and eavesdropper UAVs as a function of $P_T$ across varying horizontal distances { ($d_E^x\in25\sim 55$)} of the $E$-UAV are presented. As it can be clear from Fig. \ref{fig.sinr}(a) and \ref{fig.sinr}(b), at a distance of $d_E^x=25$ m, the UAV targets become at the same distance from the BS (see Table \ref{Tab:params}), yielding similar sensing SINR values for both targets.  However, as illustrated in Fig. \ref{fig.sinr}(a), when the $E$-UAV moves farther away, its sensing SINR decreases due to higher path loss, whereas the $L$-UAV’s sensing SINR improves as it experiences less interference from the $E$-UAV.} 

{ To provide further insights into sensing performance evaluation, in Fig. \ref{Fig:CRB}, the root CRB estimations for AoDs of the $E$-UAV are provided across varying horizontal distances of $E$-UAV, $d_E^x\in25\sim55$ m. As expected, a larger $d_E^x$ degrades root CRB estimation due to weakened sensing signal strength caused by higher path attenuation. It is also apparent from the results that an increase in $P_T$ improves the root CRB estimations. }

\begin{figure}[t]
    \centering
    \includegraphics[width=1\linewidth]{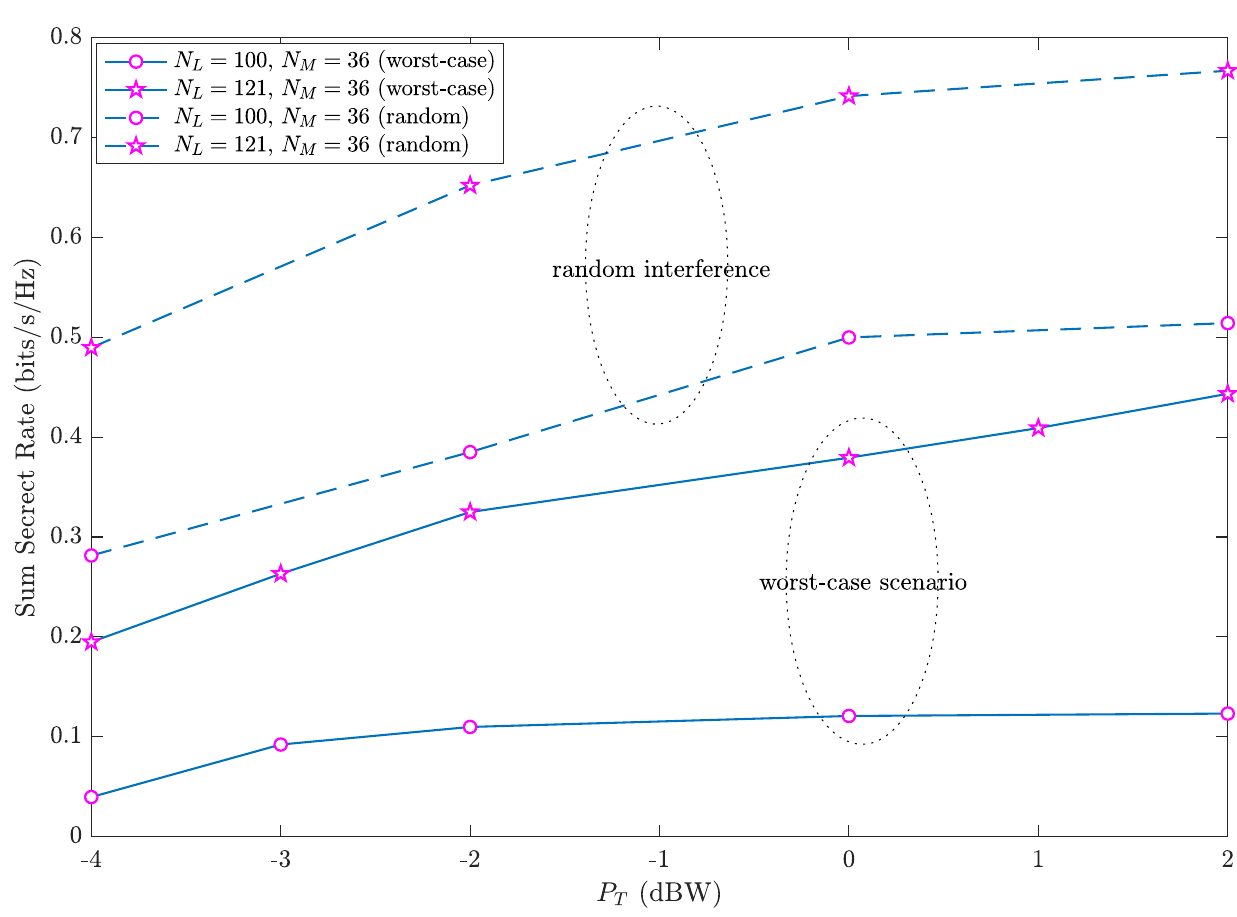}
    \caption{{ Secrecy rate performance of the worst-case malicious interference.}}
    \label{Fig:worst-case}
\end{figure}

{ In Fig. \ref{Fig:worst-case}, $S_R$ performance of the proposed scheme is illustrated for both random malicious interference and optimized interference under the worst-case scenario, where the transmit beamforming and the CSI between the BS and the $E$-UAV are compromised and perfectly known by the $E$-UAV. It is clear from the results that increasing $N_L$ improves the $S_R$ performance for both cases. Moreover, the worst-case malicious RIS interference yields a significantly lower $S_R$ than random interference, demonstrating the effectiveness of our optimization algorithm in (\ref{eq:worst}). Specifically, the optimized interference strategy exploits the compromised beamforming and CSI knowledge to intentionally degrade the $S_R$, thereby representing a stringent benchmark scenario. Nevertheless, the results indicate that a positive $S_R$ is still achievable even under worst-case conditions, confirming that reliable and secure communication remains feasible despite strong adversarial interference.}

\section{Conclusion}
In this study, a dual target-mounted RISs-aided ISAC framework has been proposed to ensure secure communication under dual security threats:  an eavesdropper target that intercepts legitimate communications and a malicious RIS that launches random interference attacks. To address these challenges,  a non-convex optimization problem is formulated to maximize the sum secrecy rate of the overall system. Then, in order to solve this problem, an SDR-based two-stage algorithm has been developed to optimize the beamforming matrix and phase shifts of the legitimate RIS. {  Moreover, a worst-case malicious interference scenario has been investigated, in which the malicious RIS interference has been optimized to maximally degrade the secrecy rate of the legitimate system, thereby evaluating the robustness of the proposed framework under the most adversarial conditions.} Furthermore, comprehensive computer simulations have been conducted to investigate the effectiveness of the proposed algorithms on the secrecy rate and sensing performance metrics across various system configurations. Future research will focus on leveraging RISs to strengthen the security of sensing signals, particularly in dynamic scenarios involving moving targets.


\bibliographystyle{IEEEtran}
\bibliography{main.bib}
\end{document}